\documentclass[%
 reprint,aip,
superscriptaddress,
 amsmath,amssymb,
]{revtex4-2}
\usepackage{graphicx}
\usepackage{dcolumn}
\usepackage{bm}
\usepackage{soul}
\usepackage[colorlinks]{hyperref}

\usepackage[utf8]{inputenc}
\usepackage[T1]{fontenc}
\usepackage{mathptmx}
\usepackage{etoolbox}
\usepackage{xcolor}
\usepackage{lineno}
\usepackage{amsmath}
\usepackage{amssymb}
\usepackage{mathtools}
\usepackage{cleveref}
\usepackage[english]{babel}
\usepackage{color}

\makeatletter
\def\@email#1#2{%
 \endgroup
 \patchcmd{\titleblock@produce}
  {\frontmatter@RRAPformat}
  {\frontmatter@RRAPformat{\produce@RRAP{*#1\href{mailto:#2}{#2}}}\frontmatter@RRAPformat}
  {}{}
}%

\makeatother
\begin{document}

\preprint{AIP/123-QED}

\title[]{Transitions between dissipative localized structures in the simplified Gilad-Meron model for dryland plant ecology}

\author{Fahad Al Saadi}
\email{fahad.alsaadi@mtc.edu.om}
\affiliation{Department of Systems Engineering, Military Technological College, Muscat, Oman}

\author{Pedro Parra-Rivas}%
 \email{pedro.parra-rivas@uinroma1.it}
\affiliation{Dipartimento di Ingegneria dell’Informazione$,$ Elettronica e Telecomunicazioni$,$
Sapienza Universit\'a di Roma$,$ via Eudossiana 18$,$ 00184 Rome}%

\date{\today}

\begin{abstract}
Spatially extended patterns and multistability of possible different states is common in many ecosystems, and their combination has an important impact on their dynamical behaviours. One potential combination involves tristability between a patterned state and two different uniform states. Using a simplified version of the Gilad-Meron model for dryland ecosystems, 
we study the organization, in bifurcation terms, of the localized structures arising in tristable regimes. These states are generally related with the concept of wave front locking, and appear in the form of spots and gaps of vegetation. We find that the coexistence of localized spots and gaps, within tristable configurations, yield the appearance of hybrid states. We also study the emergence of spatiotemporal localized states consisting in a portion of a periodic pattern embedded in a uniform Hopf-like oscillatory background in a subcritical Turing-Hopf dynamical regime. 

\end{abstract}

\maketitle

\begin{quotation}
The appearance of localized and extended structures in systems far from the thermodynamical equilibrium has sparked much theoretical research. These states normally appear in multistable regimes where different stable states coexist for the same range of parameters. Most of the studies have focused on bistable configurations, where either the two coexisting states are uniform, or one of them is uniform while the other a patterned state. These two configurations are common in fields of nonlinear research, ranging from nonlinear optics to plant ecology. In the latter, different tristable regimes have been identified, although a complete understanding of the bifurcation structure of the localized states appearing in such configurations is still needed. Here we study theoretically the transition between such configurations in a dryland ecosystem context, unveiling their formation mechanisms and spatiotemporal dynamics. 
\end{quotation}

\section{Introduction}

The emergence of spatial localization is one of the most interesting phenomena occurring in non-equilibrium systems \cite{cross_pattern_1993,cross_pattern_2009}. Thermodynamically, these are open systems where there is a continuous exchange of energy and matter with the surrounding media. In this context, localization occurs when a four-fold balance condition is satisfied: spatial coupling (e.g., diffusion) counterbalance nonlinearity, whereas dissipation is counteracted by external forcing \cite{akhmediev_dissipative_2008}. These localized states are commonly known as dissipative localized structures, hereafter LSs. LSs have been reported in different physical systems ranging from nonlinear optics and ferromagnetic fluids to atmospheric phenomena, biology and ecology \cite{akhmediev_dissipative_2008,descalzi_localized_2011,knobloch_spatial_2015, Fahadpray}. 

The formation of spatially LSs in  dissipative systems is intrinsically related with the presence of multistability between different stable extended states, which may be spatially uniform or not. The most common configuration involves the coexistence of two stable states, and is referred to as bistability. The two bistable states can be either (i) two uniform states, or (ii) a uniform and patterned state. If such coexistence is provided, wave fronts connecting the two states can form, and under suitable conditions may lock leading to the formation of LSs of different extensions \cite{coullet_nature_1987,thual_localized_1988,coullet_localized_2002}. Bistable scenarios (i) and (ii) can also occur simultaneous leading to a tristable configuration, whose potential dynamical implications are yet to be fully understood. 

 LSs have been reported in different ecosystems such as drylands
\cite{macfadyen_vegetation_1950,becker_fairy_2000,van_rooyen_mysterious_2004,meron_localized_2007,VegBxl2011Deblauwe,meron_pattern-formation_2012,tschinkel_experiments_2015,getzin_discovery_2016} and marine sea-grass \cite{ruiz-reynes_fairy_2017} ecosystems. In arid and semiarid regions, LSs can appear as spots \cite{lejeune_localized_2002, escaff_localized_2015, zelnik_yuval_r._regime_2013},
gaps \cite{tlidi_vegetation_2008, Veg_FC_2014_Fernandez-Oto, zelnik_localized_2016}, and rings \cite{VegIsrael2007Sheffer2007, VegIsrael2011Sheffer,VegIsrael2019YizhaqRings}, among others. A very important problem that these ecosystem face is the occurrence of desertification 
which can take place through the slow advance of the barren state (i.e. front propagation) \cite{zelnik_desertification_2017}, or abrupt collapse \cite{VegOtros2012Scheffer,VegIsrael2012Bel}. LSs generally appear close to the desertification onset \cite{franklin_organizing_2020,parra-rivas_formation_2020}, and therefore their study is key to understand such a process. Tristability has been recently found in different reaction-diffusion systems, including models describing dryland ecosystems \cite{zelnik_gradual_2015,getzin_discovery_2016,zelnik_wavelength_2017,zelnik_desertification_2017}. The possible implications of tristability for regime shifts (e.g., desertification) in dryland ecosystems has been discussed in Ref.\cite{zelnik_desertification_2017}

In this paper we present a detailed analysis of the bifurcation structure and stability of spatially localized states appearing when crossing from scenario (i) to (ii), in the context of plant ecology.
Here we provide a classification of the different states and bifurcation structures which may appear in this type of systems, their modification when crossing different multitable regimes, and the dynamics that they may encounter. The starting point in our study is the dryland vegetation model introduced by Gilad {\it et al.} \cite{gilad2004ecosystem}, which captures quite well
a wide variety of vegetation pattern-formation phenomena. In what follows, we refer to this model as the Gilad-Meron model, hereafter GM model. 

This paper is organized as follows. In Sec.~\ref{sec:1} we introduce the GM model and its context of applicability: plant ecology in semi-arid regions. In Sec.~\ref{sec:2} we present some preliminary results regarding the homogeneous steady states (HSSs) of the system and their linear stability against uniform and non-uniform perturbations. Moreover, we present the main stability and dynamical regions, identifying those of bi- and tristability. Furthermore, we present a spatial dynamics view of the stationary problem. Section~\ref{sec:3} is the main core of the paper where we first show the complete phase diagram of the system and analyze the modification of the bifurcation structure associated with spatially periodic pattern and LSs. In Sec.~\ref{sec:4} we study the spatio-temporal dynamics of LSs in a Turing-Hopf dynamical regime. Finally, Sec.~\ref{sec:5} presents a general discussion of our work and the main conclusions of the paper.

\section{The Gilad-Meron model}\label{sec:1}
We consider a simplified version of the GM
model, relevant to sandy soil for which overland water flow
is insignificant \cite{zelnik_gradual_2015}. The simplified model consists of two
state variables: the areal densities of the above-ground
vegetation biomass $b({\bf r},t)$ and of the soil-water content $w({\bf r},t)$. Expressed in terms of adimensional state variables and parameters, the model reads \cite{meron2016pattern}
\begin{subequations}\label{mode1}
	\begin{align}
	  \frac{\partial b}{\partial t}&= b w \left(1+\eta b \right)^{2}\left(1-b\right)-b+\delta^2  \nabla^2b,  \label{veg1} \\
	\frac{\partial w}{\partial t}&=p-\displaystyle{\frac{n w}{1+\rho b}}-\alpha b w \left(1+\eta b\right)^{2}+\nabla^2 w, \label{Veg2} \quad
	\end{align}	
\end{subequations}
The control parameters are 
 $\eta$ (root-to-shoot ratio), $p$ (precipitation rate), $n$ (evaporation rate), $\alpha$ (water uptake) and $\rho$ (reduced evaporation), and are all positive. $\nabla^2\equiv\partial_x^2+\partial_y^2$ represents the water and seed diffusion, where $\delta$ is the ratio of the previous diffusion coefficients. In this work we focus on a 1D problem and we take $\nabla^2=\partial_x^2$. Detailed information regarding the choice of dimensionless variables within the system \eqref{veg}, as well as possible implications, can be found in \cite{meron2016pattern,zelnik2015gradual,fernandez2019front}. In this model we define the diffusion coefficient in front of $\nabla^2 b$, in contrast to the standard version of the model. In the following, we fix $\delta=0.36$, $\rho=1$, $\alpha=0.5$ and $\eta=3.5$, and use as control parameters the precipitation and evaporation rates $p$ and $n$, respectively\cite{fernandez2019front}.

\section{Preliminaries: Linear stability of the homogeneous steady state, spatial dynamics, and localization}\label{sec:2}

\subsection{Homogeneous steady state}
The first step in order to analyze Eq.~(\ref{mode1}) is to study the HSS solution and its linear stability analysis. The HSS is determined by setting both the time and space derivatives to zero in Eq.~(\ref{mode1}). By doing so we obtain the system 
\begin{subequations}\label{veg}
	\begin{align}
	   b w \left(1+\eta b \right)^{2}\left(1-b\right)-b=0, \\
p-\displaystyle{\frac{n w}{1+\rho b}}-\alpha b w \left(1+\eta b\right)^{2}=0\quad,
	\end{align}	
\end{subequations}
which has two solutions.
\begin{figure}[!t]
\centering
\includegraphics[scale=0.8]{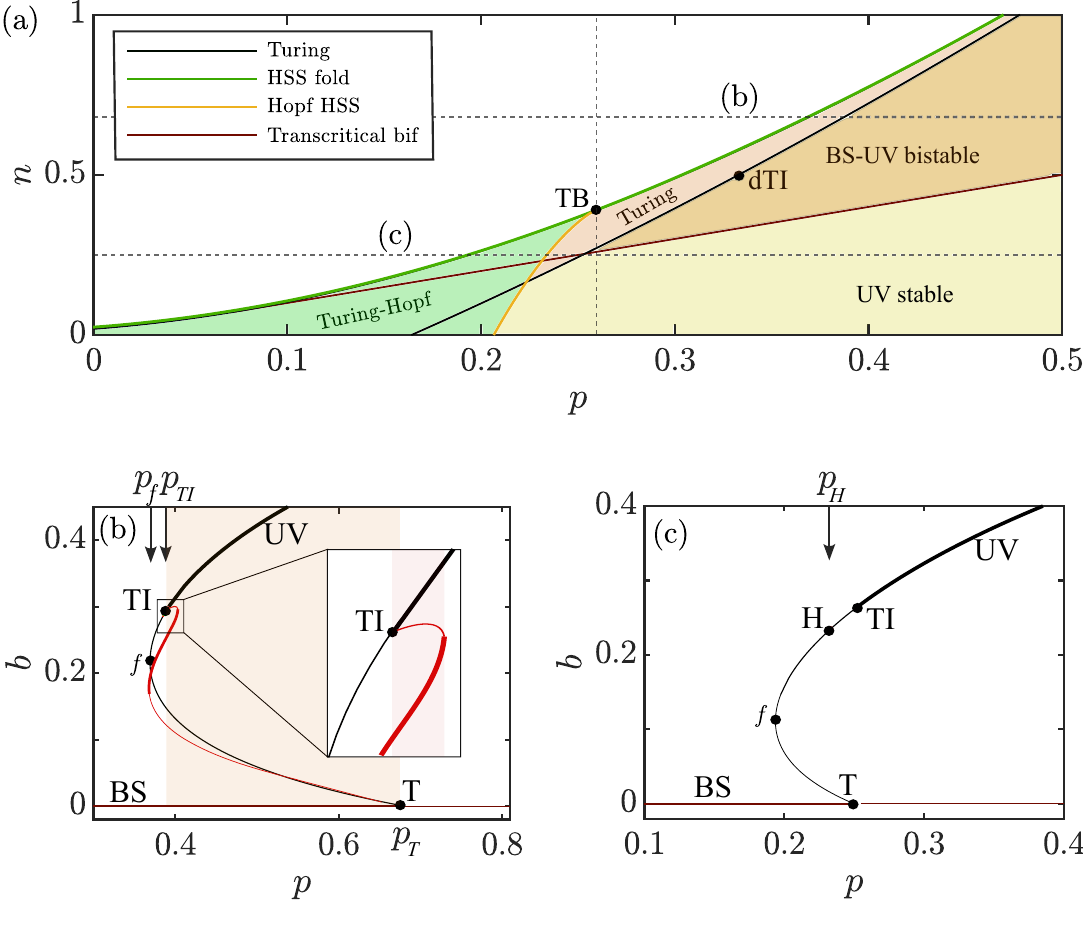}
\includegraphics[scale=1]{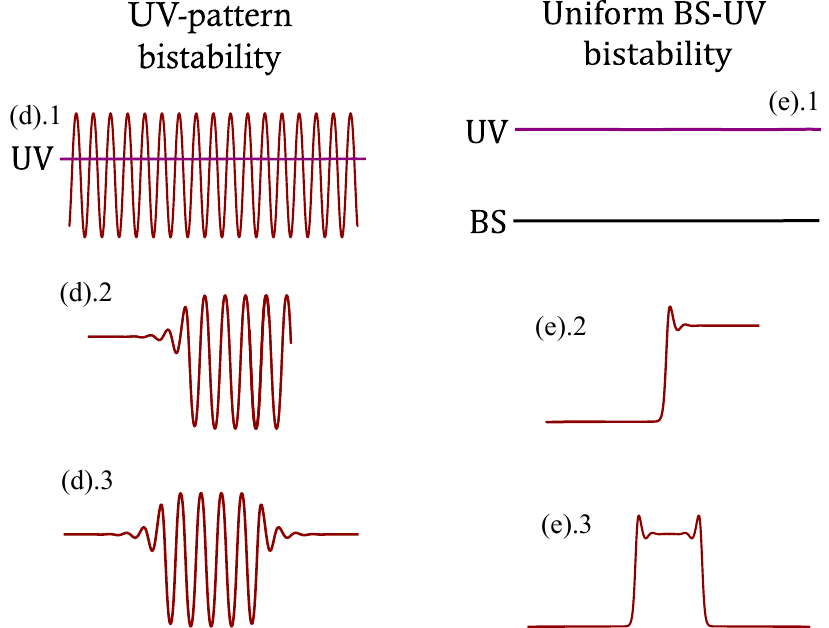}
\caption{(a) Phase-diagram in the $(p,n)$-parameter space showing the main bifurcations of the HSS of the system. Panel (b,c) show the bifurcation diagrams of the HSSs and its linear stability respect to homogeneous perturbations $(k=0)$ for $n=0.68$, and $n=0.25$, respectively. The UV state undergoes a transcritical, fold and Hopf bifurcations located at $p=p_T$, $p=p_f$, and $p=p_H$ respectively. In both cases, TI appears at $p=p_{TI}$. (d) shows the UV-pattern bistability, and (e) the BS-UV uniform bistability. Thick (thin) lines correspond to stable (unstable) states. 
}
\label{Fig1}
\end{figure}
 The simplest one is the bare soil (BS) state $\left(b,w\right) = \left(0,p/n\right)$ which corresponds to the absence of vegetation. The other HSS corresponds to uniform vegetation (UV) $(b_u,w_u)$, where 
\begin{equation}
 w_u=-{\frac {1}{ \left( b_s-1 \right)  \left( \eta\,b_s+1 \right) ^{2}}},  
\end{equation}
and $b_u$ satisfies the following quartic equation 
\begin{equation*}
    c_4b_u^4+c_3b_u^3+c_2b_u^2+c_1b_u+c_0=0,
\end{equation*}
with the coefficients 
\begin{equation*}
    c_4\equiv{\eta}^{2}\rho\left( p+\alpha \right),
\end{equation*}
\begin{equation*}
    c_3\equiv -  \left[
 \left( \rho-1 \right) p-\alpha \right] \eta^2-2\rho \left( p+\alpha
 \right)\eta,
\end{equation*}
\begin{equation*}
    c_2\equiv -p{\eta}^{2}+ \left[  \left( -
2\,\rho+2 \right) p+2\,\alpha \right] \eta+\rho\, \left( p+\alpha\right),
\end{equation*}
\begin{align*}
    c_1\equiv -2\,p\eta+ \left( -\rho+1 \right) p+\alpha, && c_0\equiv n-p. 
\end{align*}

The dependence of BS and UV with $p$ is depicted in Fig.~\ref{Fig1}(b) for $n=0.68$. The BS state undergoes a transcritical (T) bifurcation at $p=p_T\equiv 0.68 $. From $p_T$ the UV state arise subcritically, and undergoes a fold bifurcation $f$ at $p=p_f\equiv 0.1937$.
The dependence of the transcritical and fold bifurcations with $n$ is depicted in the two-parameter bifurcation diagram plotted in Fig.~\ref{Fig1}(a). By decreasing $n$, so does the subcriticality, as the fold and transcritical bifurcation approach one another. Such a situation is depicted in Fig.~\ref{Fig1}(c) for $n=0.25$.

\subsection{Linear stability analysis}
A more detailed study of these states requires a linear stability analysis. The basic idea behind this analysis is to determine how the HSS responds against modulated perturbations of the form ${\rm cos}(\sigma t+ikx)$, or what is equivalent, to study weakly modulated solutions of the form     
\begin{equation}
(b,w)^T=(b_h,w_h)^T+(\xi_1,\xi_2)^T e^{ikx+\sigma t}+c.c.,
\label{eq:eigs}
\end{equation}
where $|\xi_m|\ll1$ for $m=1,2$, $k$ is the spatial wave-number of the modulation and $\sigma$ is the growth rate of the perturbation. 
By substituting \eqref{eq:eigs} into \eqref{mode1}, and linearizing it around any of the HSS we obtain the linear system 
\begin{equation*}
    \partial_t\xi=J \xi
\end{equation*}
where $\mathbf{\xi}\equiv(\xi_1,\xi_2)$ and  $J$ is the Jacobian linear operator associated with Eq.~(\ref{mode1}), once evaluated at $(b_h,w_h)$.
This equation has non-trivial solutions if the solvability condition 
\begin{equation} \label{des}
\det\left(J-k^2 D-\sigma I \right)=0,
\end{equation}
is satisfied,
where
$D$ is the diffusion coefficients matrix $D\equiv\mbox{diag}(1,\delta^2)$ and $I$ is
the identity matrix.

\begin{figure*}
\centering
\includegraphics[scale=1]{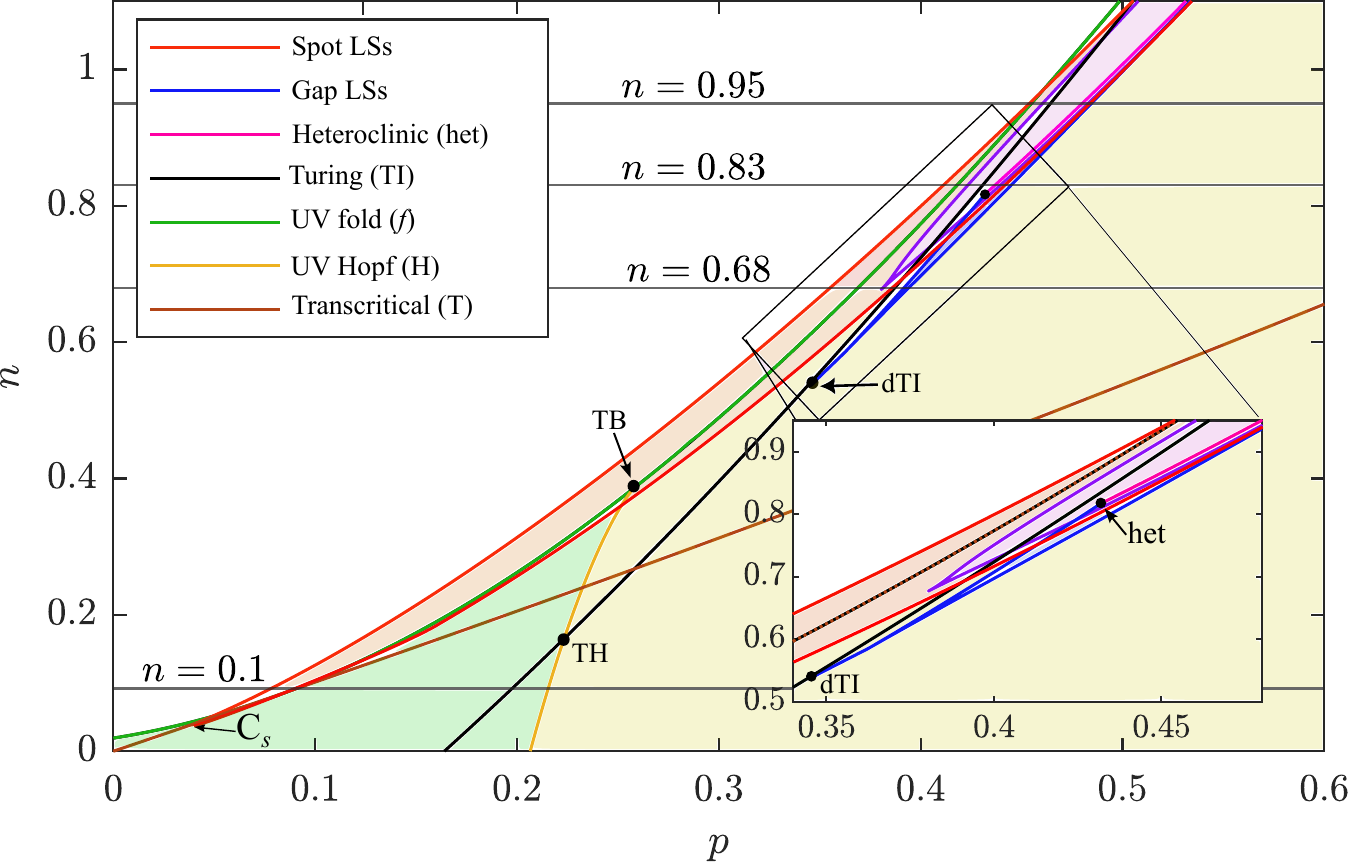}
\caption{Two-dimensional phase diagram in the $(p,n)$-parameter space  for $\alpha=0.5$, $\rho=1$, $\eta=3.5$, $\delta=0.36$. This diagram depicts the main bifurcation lines of the system, and the different stability and dynamical regions. The horizontal lines correspond to the bifurcation diagrams shown in Figs.~\ref{HomSnakn1}, \ref{HomSnakn68}, \ref{HomSnakn83}, and \ref{Collapsed} for $n=0.1,0.68,0.83,$ and $0.95$ respectively.}
\label{phase_diagram}
\end{figure*}
The HSS is stable if Re$[\sigma]$<0 and unstable otherwise. The transition between these two situations occurs when Re[$\sigma$]=0 where a bifurcation or instability takes place. In our context, three different bifurcations may arise:
\begin{itemize}
\item If Im$[\sigma(k)]=$Re$[\sigma(k)]=0$ at $k=0$, the flat solution undergoes stationary bifurcations. To this type correspond the transcritilcal and fold bifurcations previously introduced. 
 \item If Im$[\sigma(k)]\neq0$ and Re$[\sigma(k)]=0$ at $k=0$, the flat solution undergoes a Hopf bifurcation (HB).
 \item If Im$[\sigma(k)]=0$, Re$[\sigma(k)]=0$, and $d{\rm Re}[\sigma]/dk=0$ at $k=k_T$, the HSS undergoes a Turing instability (TI).
\end{itemize}

 Analytical expressions describing the onsets and features of these instabilities are too intricate and here we only use its graphical representation. 
The dependence of the position of these bifurcations in the parameter plane $(p,n)$ is depicted in Fig.~\ref{Fig1}(a) together with the main stable and unstable regions. In Figs.~\ref{Fig1}(b) and (c) the HSSs stability is illustrated for the BS and UV states, where solid thick (thin) lines correspond to stable (unstable) states. In both cases, BS is stable for $p<p_T$, and unstable otherwise. In Fig.~\ref{Fig1}(b) [$n=0.68$], the UV state undergoes a TI at $p_{TI}$ and becomes unstable to modulated states within the interval $p_f<p<p_{TI}$. Decreasing $n$, the HB emerges from 
the fold [see black dot in Fig.~\ref{Fig1}(a)], and eventually crosses with the TI in a codimension-two Turing-Hopf (TH) bifurcation. This point leads to a complex type of dynamics characterized by the competition between the previous temporal and spatial symmetry breaking instabilities \cite{meixner_generic_1997,just_spatiotemporal_2001}. That situation is illustrated in Fig.~\ref{Fig1}(c) for $n=0.25$. Here, the HB occurs after TI, and two dynamical regions appear. For $p_H<p<p_{TI}$ UV is Turing unstable (i.e., unstable against modulated states with finite $k$). For $p_f<p<p_H$, the Turing and Hopf modes may interact, leading much complex dynamics. This region corresponds to the green region depicted in Fig.~\ref{Fig1}(a).

\subsection{Spatial dynamics framework}
In this work we focus on the emergence of steady states (i.e., time-independent states satisfying $\partial_tb=\partial_t w=0$) of Eq.~(\ref{veg}), and therefore in solutions of the stationary system
\begin{subequations}\label{staveg}
	\begin{align}
	  \delta^2  b''(x)&= -b w \left(1+\eta b \right)^{2}\left(1-b\right)+b,  \label{veg1} \\
w''(x)&=-p+\displaystyle{\frac{n w}{1+\rho b}}+\alpha b w \left(1+\eta b\right)^{2}. \label{staVeg2} \quad
	\end{align}	
\end{subequations}

By defining the new variables $\left[u_1(x),u_2(x),u_3(x),u_4(x)\right]\equiv\left[b(x),w(x),b'(x),w'(x)\right]$, the previous system can be recast into the 4D dynamical system 
\begin{equation}\label{DS}
  \begin{array}{ll}
  u'_1(x)=& u_2(x)\\\\
  u'_2(x)=& -u_1 u_3 \left(1+\eta u_1 \right)^{2}\left(1-u_1\right)+u_1\\\\
  u'_3(x)=& u_4(x)\\\\
  u'_4(x)=& -p+\displaystyle{\frac{n u_3}{1+\rho u_1}}+\alpha u_1 u_3 \left(1+\eta u_1\right)^{2}\\\\
  \end{array}  
\end{equation}
With this reformulation of the problem, known as {\it spatial dynamics}, we can establish a duality between the time-independent states of the system (HSSs, periodic patterns and LSs) and the solutions of Eq.~(\ref{DS}). In this context, the HSS of Eq.~(\ref{staveg}) corresponds to a fixed point of Eq.~(\ref{DS}), a spatially modulated state to a limit cycle, and wave fronts and LSs to heteroclinic and homoclinic orbits, respectively. 

In this framework, we apply 
path-continuation methods \cite{allgower_numerical_1990,krauskopf_numerical_2007,uecker_numerical_2021} to compute the bifurcation structure of extended and LSs. To do so we solve the boundary value problem associated with Eq.~(\ref{DS}) considering Neuman boundary conditions in half of the domain $[0,L/2]$, and using the open distribution software AUTO-07p for numerical continuation \cite{doedel_auto-07p_2007}. In what follows we fix $L=100$.

The temporal stability of the different steady states is then computed from the eigenvalue problem obtained after linearizing Eq.~(\ref{mode1}) around each steady state.

\subsection{Spatial localization: bistability and front locking}
The formation of a wide variety of LSs in dissipative systems is generally related with the locking of wave fronts connecting two different, but coexisting, stable states \cite{colet_formation_2014}.
Depending on the nature of the two coexisting states, two main scenarios are commonly considered : either the two states are uniform, or one of them is uniform while the other one consists in a spatially modulated state. In the first case we may talk about {\it uniform bistability}, whereas the second can be referred to as {\it pattern-uniform bistability}. 

Here, pattern-uniform bistability appears between the UV and spatially subcritical modulated (periodic pattern) state [see Fig.~\ref{Fig1}(d).1]. The periodic pattern bifurcates subcritically from the TI [see the periodic pattern red diagram in the close-up view of Fig.~\ref{Fig1}(b)] once the codim-2 degenerate TI point dTI is crossed [see Fig.~\ref{Fig1}(a)].    Below this point, the pattern is supercritical, while above is subcritical.   The wave fronts connecting UV with the subcritical pattern are like the one depicted in Fig.~\ref{Fig1}(d).2. As illustrated in Fig.~\ref{Fig1}(d).3, their locking leads to the formation of LSs consisting on a portion of the pattern embedded in the uniform state \cite{woods_heteroclinic_1999,coullet_stable_2000,makrides_existence_2019}. These type of states undergo a bifurcation structure known as {\it standard homoclinic snaking}, whose morphology is a direct consequence of the wave front interaction \cite{woods_heteroclinic_1999}. 

In the uniform scenario of our model the two coexisting states are the trivial BS state and the UV one [see Fig.~\ref{Fig1}(e).1]. This bistable region is located in-between T and TI [see Fig.~\ref{Fig1}(b) and the red shadowed area in Fig.~\ref{Fig1}(a)]. The wave-fronts connecting these states are shown in Fig.~\ref{Fig1}(e).2. While the front approaches monotonically BS, it reaches UV in a damped oscillatory fashion. The presence of these "oscillatory tails" allows the locking of two fronts of different polarity and the formation of LSs like the one shown in Fig.~\ref{Fig1}(e).3 \cite{coullet_localized_2002,knobloch_homoclinic_2005}. These states consist in a plateau formed by a portion of UV embedded in BS, and undergo a bifurcation structure known as {\it collapse homoclinic snaking} \cite{knobloch_homoclinic_2005}.

Varying the control parameters of the system we can find regimes where both type of bistability may coexist, leading to a tristability regime, and to the formation of many different types of hybrid states. The implication of tristability on pattern forming ecosystems have been studied in Ref.~\cite{zelnik_implications_2018}, and we have recently analyse the transition between standard and collapsed homoclinic snaking in such regime in a prototypical pattern forming model: the Swift-Hohenberg equation \cite{parra-rivas_organization_2022}. In the following section we perform a detailed bifurcation analysis of our system, unveiling the organization of the LSs within the tristable regime.

\section{Bifurcation structure for gaps and spots}\label{sec:3}
The $(p,n)$-bifurcation diagram depicted in Fig.~\ref{phase_diagram} shows a more complete version of Fig.~\ref{Fig1}(a), where together with 
the main bifurcation lines and stability regions for HSSs, we have added those corresponding to spatially periodic patterns and LSs.

In this section we focus on the bifurcation structure and stability of LSs which are not affected by the Hopf instability or by the TH type of dynamics. Later, in Sec.~\ref{sec:4} we will explore the influence of such oscillatory dynamics on the LSs stability.  


\begin{figure*}[!t]
\centering
\includegraphics[scale=1]{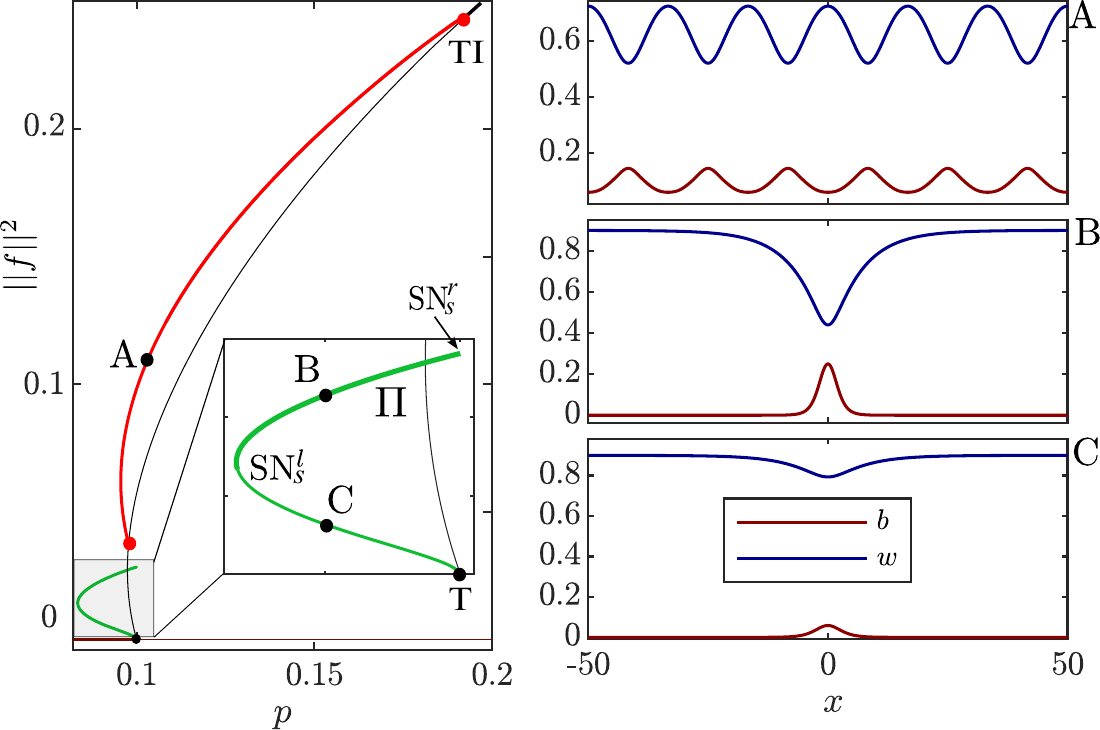}
\caption{Bifurcation diagram for $n=0.1$ and their associated LSs. The BS and UV state are plotted using solid black lines. The supercritical spatially periodic pattern state bifurcating from TI is plotted in red. Thin and thick solid lines represent unstable and stable states respectively. }
\label{HomSnakn1}
\end{figure*}
A standard way to understand the $(p,n)$-bifurcation diagram in Fig.~\ref{phase_diagram} consists in slicing it for a fixed value of either $p$ or $n$, and studying the resulting 1D bifurcation diagrams. 
In this work we fix $n$ at four different, and representative, values $n=0.1,0.68,0.83$, and $0.95$. These values correspond to the horizontal lines plotted in Fig.~\ref{phase_diagram}.
\subsection{Bifurcation diagram for $n=0.1$}\label{sec:3A}

The bifurcation structure for $n=0.1$ is depicted in Fig.~\ref{HomSnakn1}, where we plot the $L_2$-norm of the field $f=(b,w)$ composed by the biomass $b$ and soil water $w$ components, namely 
\begin{equation*}
   ||f||^2\equiv L^{-1}\int_{-L/2}^{L/2}\left(b^2(x)+w(x)^2\right)dx
\end{equation*}
as a function of $p$. This diagram shows different families of states. The solution branches in black correspond to UV already shown in Fig.~\ref{Fig1}, and BS is plotted in scarlet red. The UV state becomes unstable at the TI to spatially modulated states, which emerge supercritically (see red solid line) with a wavelength $2\pi/k_T$. An example of such state is depicted in Fig.~\ref{HomSnakn1}A. Decreasing $p$, so does the amplitude of the modulated state which eventually connects back supercritically to UV in a branching point taking place at $p\approx0.1$.

Close to the transcritical bifurcation, small amplitude LSs are captured by the asymptotic solution 
\begin{equation}\label{sol_trasncri}
    \left(\begin{array}{c} b\\w \end{array}\right)-\left(\begin{array}{c} b_{\rm BS}\\w_{\rm BS} \end{array}\right)\propto\frac{3}{2}\frac{\Delta p}{c_3}{\rm sech}^2\left(\frac{1}{2}\sqrt{-\frac{\Delta p}{c_1}}\right),
\end{equation}
where $\Delta p=p_T-p$, and the coefficients $c_{1,3}$ depend on the control parameters of the system \cite{parra-rivas_formation_2020}.
These states are unstable, and look like the one depicted in Fig.~\ref{HomSnakn1}C, where the content of soil water is a hole, while the density of biomass is a spot (i.e., spike) LS. By modifying $p$, they undergo the green bifurcation curve $\Pi$ shown at the bottom of Fig.~\ref{HomSnakn1}.

By decreasing $p$, that state undergoes the saddle-node bifurcation SN$_s^l$ where it stabilises.
The stable biomass spike state extends until SN$_s^r$, and looks like the one depicted in  
Fig.~\ref{HomSnakn1}B. SN$_s^{l,r}$ are depicted using solid red lines in Fig.~\ref{phase_diagram}. $\Pi$ extends below the bistable region between BS and the modulated states, and therefore, it existence is not related with the wave front locking mechanism. Note that the Hopf unstable UV does not affect these type of states as they rest on the BS state, which is stable until T. 

\subsection{Bifurcation diagram for $n=0.68$}\label{sec:3B}
\begin{figure*}[!t]
\centering
\includegraphics[scale=1.3]{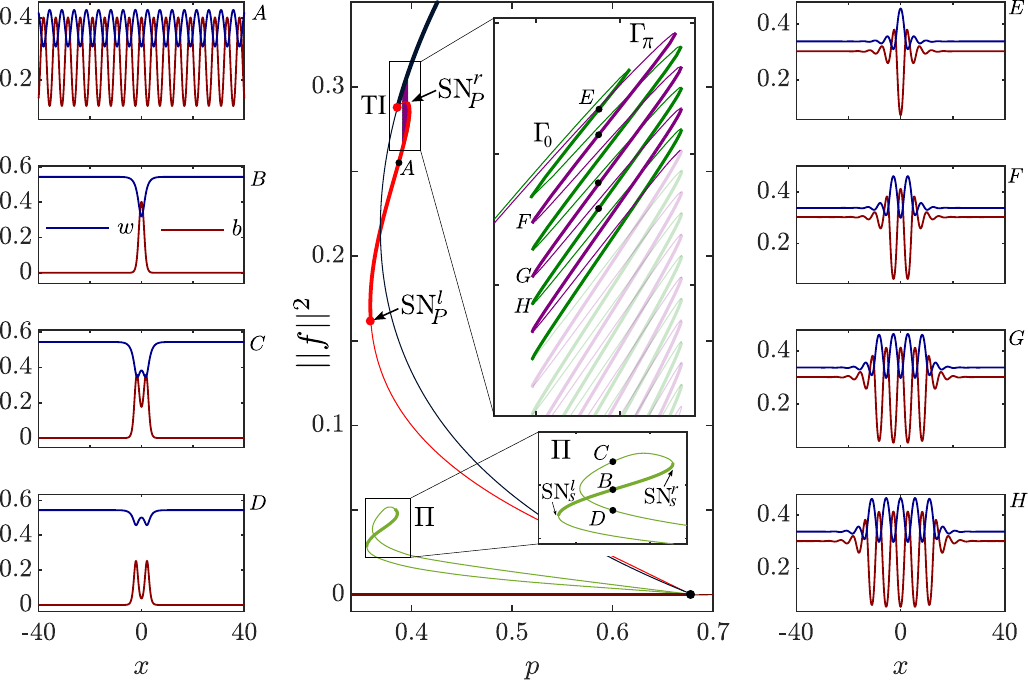}
\caption{Bifurcation diagram for $n=0.68$ and their associated LSs. The BS and UV state are plotted using solid black lines. The subcritical spatially periodic pattern state arising from TI is plotted in red. Thin and thick solid lines represent unstable and stable states respectively. Panel A is a stable spatially periodic pattern. Panels B-D correspond to spots of biomass $b$ located on the green solution branches at the bottom of the central panel. Panels E-H shows the solution profiles of localized gaps of biomass organized on a standard homoclinic snaking structure (see close-up view of the central panel). }
\label{HomSnakn68}
\end{figure*}

The main morphological modification of the bifurcation structure shown in Fig.~\ref{HomSnakn1} occurs when crossing the degenerate codim-2 Turing instability dTI (see Fig.~\ref{phase_diagram}). At this point, the modulated periodic state becomes subcritical, leading to UV-pattern bistability [see Fig.~\ref{Fig1}(d)]. 
An example of that configuration is shown in Fig.~\ref{HomSnakn68} for $n=0.68$. The spatially modulated pattern state emerges subcritically, and thus unstably, from TI at UV (see red curve in Fig.~\ref{HomSnakn68}).
 This pattern soon stabilizes at SN$_P^r$, and remains stable with decreasing $p$ until reaching SN$_P^l$. Figure~\ref{HomSnakn68}A shows an example of such a state. Once SN$_P^l$ is passed, the pattern becomes unstable again and decreases its amplitude as following down its bifurcation curve. Eventually, this state connects back to UV in a branching point taking place at the unstable UV branch near $p_T$. 

The system exhibits bistablity between UV and the spatially periodic pattern in the $p$-interval spanning from TI to SN$_P^r$. Within this interval, LSs as those depicted in Figs.~\ref{HomSnakn68}E-H  consisting in a slug of the periodic pattern embedded on the UV state arise. These LSs consist in gap-type localized patterns for $b$ and spot-like states for $w$. In bifurcation terms, they organize in a standard homoclinic snaking structure consisting in a back-and-forth oscillation of the LSs solution curves within a $p$-interval known as the snaking or pinning region \cite{woods_heteroclinic_1999,burke_homoclinic_2007,burke_snakes_2007}. Within this interval, a large multiplicity of LSs of different extensions coexist.   

Regarding $b$-profiles of the LSs we can differentiate two families of solutions: one having a minimum at $x=0$ [see Figs.~\ref{HomSnakn68}E,H], and another one where a maximum is located at $x=0$ [see Figs.~\ref{HomSnakn68}F,G]. In the following, we refer to these families using $\Gamma_0$ and $\Gamma_\pi$, respectively. Both families emerge subcritically from TI together with the periodic pattern, and very close to such bifurcation they can be described by the approximate solution 
\begin{equation}
    \left(\begin{array}{c} b\\w \end{array}\right)-\left(\begin{array}{c} b_{\rm UV}\\w_{\rm UV} \end{array}\right)\propto\sqrt{\frac{2\Delta p}{c_3}}{\rm sech}\left(\sqrt{\frac{\Delta p}{c_1}}\right){\rm cos}(k_T x+\varphi),
\end{equation}
where $\Delta p=p-p_{TI}$, $c_{1,3}$ depend on the control parameters of the system, and $\varphi=0,\pi$ correspond to the solution families $\Gamma_0$ and $\Gamma_\pi$, respectively \cite{knobloch_spatial_2015,parra-rivas_formation_2020}. The computation of this solution in the current model is beyond the scope of this paper.

Decreasing $||f||^2$, the LSs nucleate a pair of pattern rolls or peaks at either side of the center structure, increasing their width. In a infinite domain, the nucleation process never stops. However, in a finite domain, this process eventually terminates, and  $\Gamma_0$ and $\Gamma_\pi$ reconnect with two different pattern bifurcation curves \cite{parra-rivas_bifurcation_2018}.

\begin{figure*}[!t]
\centering
\includegraphics[scale=1]{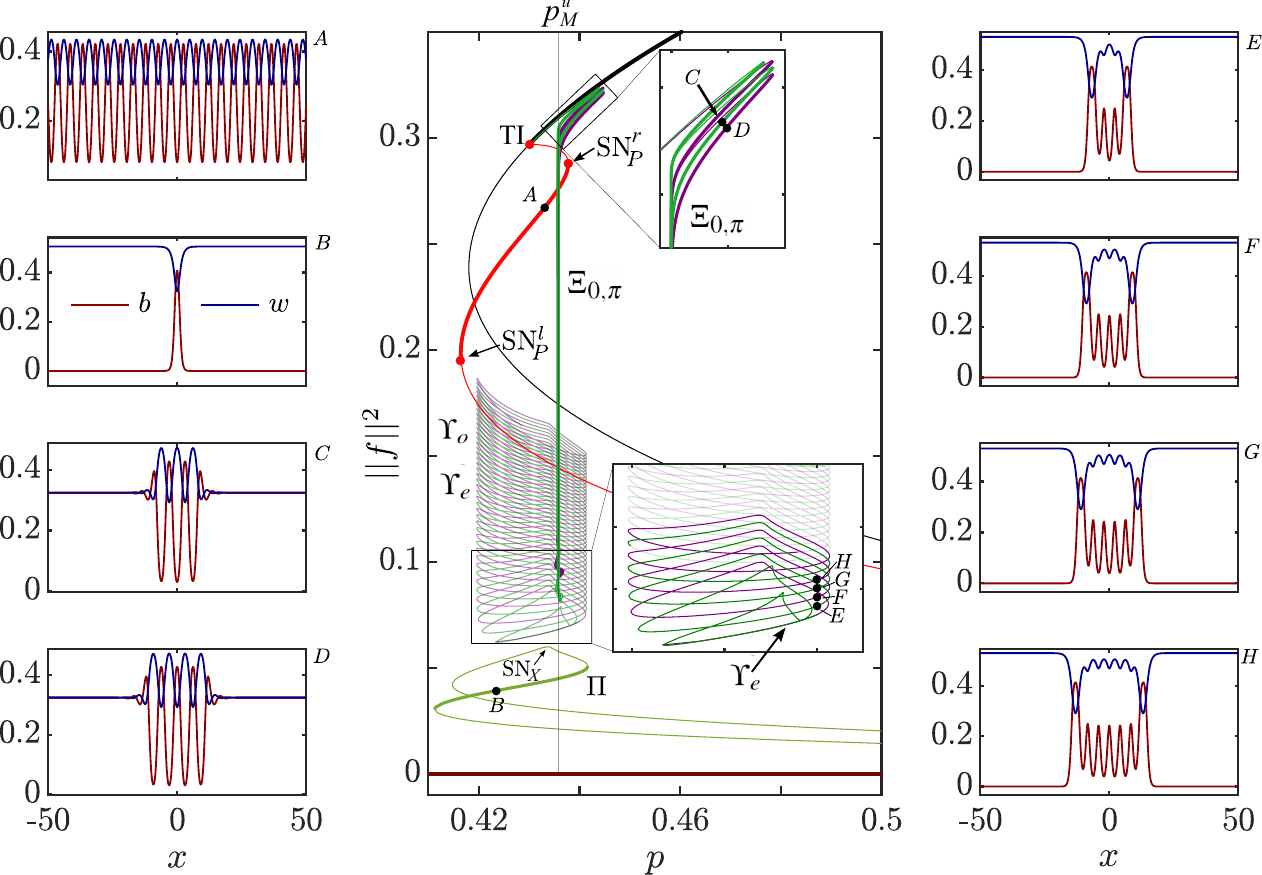}
\caption{Bifurcation diagram for $n=0.83$ and their associated LSs. The BS, UV and pattern states are illustrated similarly than in Fig.~\ref{HomSnakn83}, where thin and thick solid lines represent unstable and stable states respectively. Panel A is a stable spatially periodic pattern. Panel B corresponds to the single-peak spot of biomass located on $\Pi$. Panels C-D correspond to patterned gaps of biomass $b$ which belongs to the structure $\Xi$. Panels E-H shows the solution profiles of a new kind of localized spots of biomass which form the isola $\Upsilon$. }
\label{HomSnakn83}
\end{figure*}
The $\Pi$ bifurcation curve emerging from the T bifurcation at BS [see Fig.~\ref{HomSnakn1}] persists for this value of $n$, and connects with another curve of unstable states (see Fig.~\ref{HomSnakn68}).
By decreasing $p$, the small amplitude unstable state bifurcating from T grows and stabilizes at SN$_s^l$. These stable states are like the one depicted in Fig.~\ref{HomSnakn68}B, where the biomass density consists in a spot of vegetation. Following $\Pi$ to the right, the spot keeps increasing its amplitude until reaching SN$_s^r$ where it becomes unstable and nucleates a dip in the middle. An example of this state is plotted in Fig.~\ref{HomSnakn68}C. Decreasing $||f||^2$, this unstable state keep decreasing its amplitude and eventually disappears very close to $p_T$. The saddle-node bifurcations SN$_s^{l,r}$ are plotted in Fig.~\ref{phase_diagram} using a red solid line. Decreasing $n$, these two folds eventually meet and disappear at the cusp bifurcation $C_s$. Below this point, spot states exist no more.  
 
\begin{figure*}[!t]
\centering
\includegraphics[scale=1]{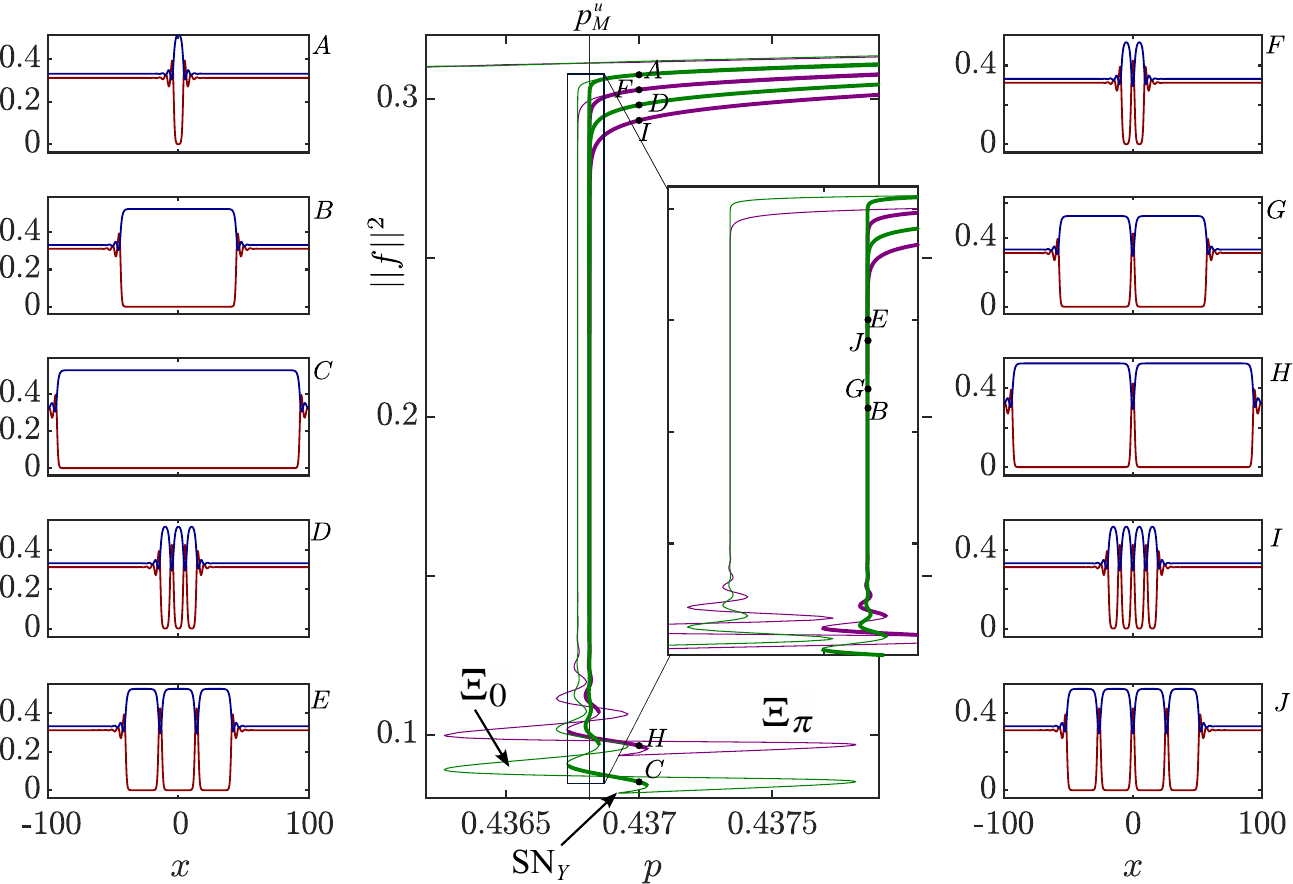}
\caption{Close-view of the bifurcation structure $\Xi_0$ and $\Xi_\pi$ shown in Fig.~\ref{HomSnakn83}, and modification of the LSs profiles along them.}
\label{Xistructure}
\end{figure*}

\begin{figure*}[!t]
\centering
\includegraphics[scale=1.3]{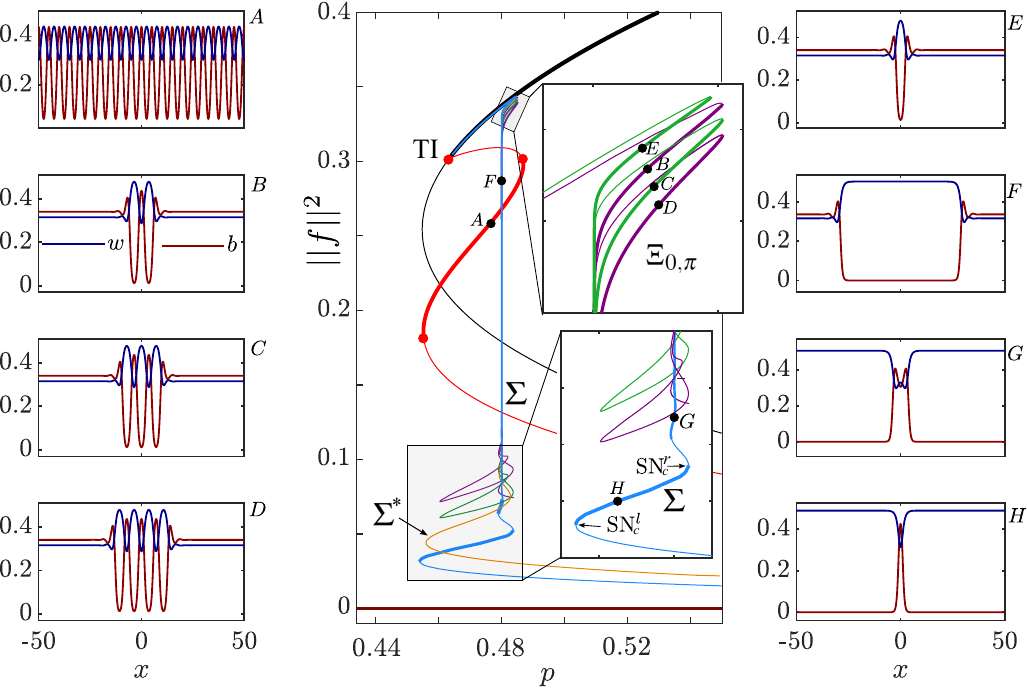}
\caption{Bifurcation diagram for $n=0.95$. The blue curve corresponds to the collapsed homoclinic snaking $\Sigma$. The green and purple curves are $\Xi_0$ and $\Xi_\pi$, respectively, and in red the solution curve for the spatially periodic pattern is plotted. Solution profiles plotted in panels B-D correspond to the gap states belonging to $\Xi_{0,\pi}$. Panels E-H show the modification of the LSs along $\Sigma$. }
\label{Collapsed}
\end{figure*}

\subsection{Bifurcation diagram for $n=0.83$}\label{sec:3C}
The bifurcation structure illustrated in Fig.~\ref{HomSnakn68} persists until reaching the codimension-two heteroclinic (het) bifurcation point (see $\bullet$ in the inset of Fig.~\ref{phase_diagram}). After crossing this point, uniform fronts (i.e., heteroclinic orbits) connecting back-and-forth BS and UV start to lock one another leading to the formation of heteroclinic cycles, i.e., LSs \cite{knobloch_homoclinic_2005}. Around this point, a transition region exists where the standard homoclinic snaking is destroyed in favor of new structural organizations\cite{zelnik_implications_2018,al_saadi_unified_2021,parra-rivas_organization_2022,ALSAADI2022}. This transition has been analyzed in detail in the context of the Swift-Hohenberg equation\cite{parra-rivas_organization_2022} and in the following we refer to this work for further details. 

This scenario is depicted in Fig.~\ref{HomSnakn83} for $n=0.83$. The spatially periodic pattern [see Fig.~\ref{HomSnakn83}A] undergoes the same structure than the one described in Fig.~\ref{HomSnakn68}. The $\Pi$ structure expands, and the biomass spot [see Fig.~\ref{HomSnakn83}B] increases its region of stability. The enlargement of this region can be appreciated in the phase diagram of Fig.~\ref{phase_diagram} from the separation of the red lines representing SN$_s^{l,r}$. One particularity of $\Pi$ is that it has developed a tip in the top unstable branch that corresponds to the saddle-node bifurcation SN$_X$. 

For this value of $n$, localized pattern persist [see Fig.~\ref{HomSnakn83}C, D], although they organize differently [see the $\Xi_{0,\pi}$ structures shown in Fig.~\ref{HomSnakn83}]. The $\Xi_{0,\pi}$ curves emerge from TI at UV and follow the uniform Maxwell point (i.e., the heteroclinic bifurcation) until low values of $||f||^2$. A detailed description of these diagrams around $p^u_M$ is shown in Fig.~\ref{Xistructure}. $\Xi_0$ (see green curve in Fig.~\ref{Xistructure}) is associated with the single peak gap state bifurcating from TI, which after growing in amplitude looks like the profile shown in Fig.~\ref{Xistructure}A. Note that this state is slightly wider that the single dip spot state depicted in Fig.~\ref{HomSnakn83}B. Proceeding down in $\Xi_0$, the width of such state increases until two well defined fronts connecting BS and UV form. An example of a very broad state of this type is plotted in Fig.~\ref{Xistructure}B. Following $\Xi_0$ further down, the two fronts reach the boundaries of the domain forming a biomass spot with three bumps [see Fig.~\ref{Xistructure}C]. This last state is stable. The lower point of $\Xi_0$ is the saddle-node bifurcation SN$_Y$. This bifurcation will be essential for the reorganization of the bifurcation scenario as $n$ increases. 

The three peak gap state shown in Fig.~\ref{HomSnakn83}C follows a similar process. Approaching $p_M^u$, the three peaks of this structure widen [see Fig.~\ref{Xistructure}D] and decreasing $||f||^2$ they become the state depicted in Fig.~\ref{Xistructure}E. This state consists in two centered single-peak spots and a very broad state formed due to the locking of uniform fronts. Decreasing even further in the diagram, the previous fronts move toward the boundaries, leading to LSs with a few bumps (not shown here). 

Along $\Xi_\pi$ the process is similar. Figures~\ref{Xistructure}F-G show the modification of a two-gap biomass state as proceeding down in $\Xi_\pi$, and Figs.~\ref{Xistructure}I-J show a similar modification of a four-gap state.  

\begin{figure*}[!t]
\centering
\includegraphics[scale=0.8]{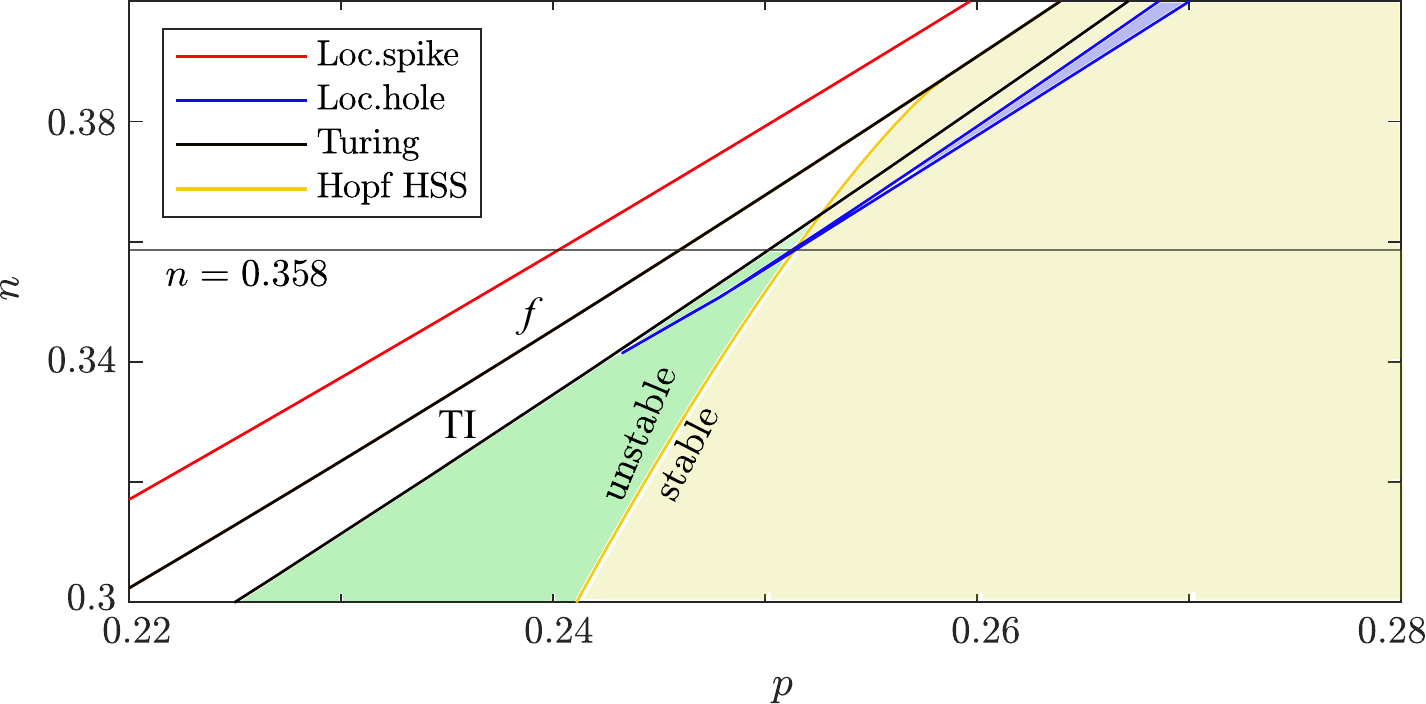}  
\caption{Two-dimensional phase diagram in the $(p,n)$-parameter space  for $\alpha=0.5$, $\rho=1$, $\eta=3.5$, $\delta=0.6$. For this set of parameters the homoclinic snaking region of gap states intersects HB.}
\label{bioep608}
\end{figure*}
Another feature of this scenario is the presence of a set of isolas, labeled $\Upsilon$, associated with the states shown in Fig.~\ref{HomSnakn83}E-H. These consist in localized spots of biomass formed by two external peaks filled by a number spatially periodic state rolls. There are two type of isolas: one with an odd number of rolls $\Upsilon_o$ (in purple) corresponding to the states depicted in Figs.~\ref{HomSnakn83}F,H, and another one with an even number of rolls $\Upsilon_e$ (in green) corresponding to Figs.~\ref{HomSnakn83}E,G. Proceeding up in these isolas, the number of rolls in between the external peaks increases by two each time that the  right folds are passed. Our linear stability analysis reveals that these states are unstable. 


\subsection{Bifurcation diagram for $n=0.95$}\label{sec:3D}
Increasing the value of $n$, a necking bifurcation N occurs at $(p,n)=(p_N,n_N)\approx(0.448,0.867)$ where SN$_X$ [see Fig.~\ref{HomSnakn83}] and SN$_Y$ [see Fig.~\ref{Xistructure}] meet transcritically\cite{parra-rivas_organization_2022}. As a result, a reconnection of branches takes place, leading to the two pairs of collapsed homoclinic snaking structures $\Sigma$ and $\Sigma^*$ depicted in Fig.~\ref{Collapsed} for $n=0.95$ (see blue and orange curves, respectively). $\Sigma$ is formed by half of $\Pi$ (containing the single-bump spot state shown in Fig.~\ref{HomSnakn68}B) and half of $\Xi_0$ (containing the profiles shown in Figs.~\ref{Xistructure}A-C), while $\Sigma^*$ appears from the reconnection of the resting parts. The collapsed snaking consists in a sinusoidal-like damped oscillation around the uniform Maxwell point of the system $p_M^u$, and is associated with the LSs formed exclusively through the locking of uniform fronts connecting BS and UV \cite{knobloch_homoclinic_2005}. Indeed, the shape of this bifurcation curve is directly related with the fronts interaction\cite{parra-rivas_formation_2020,parra-rivas_origin_2021}.

$\Sigma$ emanates from the transcritical bifurcation at $p_T$, where the LS has the form described by Eq.~(\ref{sol_trasncri}). Increasing $||f||^2$, the amplitude of such states increases and eventually stabilizes at SN$_c^l$. An example of this state is depicted in Fig.~\ref{Collapsed}H. After crossing SN$_c^r$, the state becomes unstable and nucleates a small dip at $x=0$. Once stabilized at the next fold, the biomass spot looks like the two-bump state depicted in Fig.~\ref{Collapsed}G. Proceeding up in $\Sigma$, new dips are nucleated around $x=0$, resulting in the widening of the LSs. This enlargement of the LSs width is related with the locking distance of the fronts, which increases as moving up in $\Sigma$. This is the same process previously undergone by the LSs along $\Xi_0$ (see Fig.~\ref{Xistructure}). An example of a wide LS is depicted in Fig.~\ref{Collapsed}F. The nucleation of dips, and therefore the enlargement of the LSs width, continues as increasing $||f||^2$, until the fronts separation reaches the boundaries of the spatial domain. At this point, the localized wide spot becomes the single-hole gap depicted in Fig.~\ref{Collapsed}E. This state maintains its stability until reaching SN$_c^t$. Eventually, this state dies out at the TI at $p=p_{TI}$. For this value of $n$, the structure $\Xi_\pi$ survives, as so do the biomass gap states plotted in Fig.~\ref{Collapsed}B-D.


\section{Spatiotemporal Turing-Hopf localized states}\label{sec:4}

Spatiotemporal localized solutions characterized by the coexistence of Turing and uniform Hopf states have been observed
in the vicinity of a codimension-two TH point, where both bifurcations are supercritical \cite{borckmans_localized_1995,de_wit_spatiotemporal_1996,tlidi_spatiotemporal_1997,tzou_homoclinic_2013}. These dynamical states consist in a static Turing localized pattern embedded in a oscillatory uniform background field, and undergo a kind of snaking structure similar to the standard homoclinic snaking of stationary states, as shown by Tzou {\it et al.} \cite{tzou_homoclinic_2013}. In that work, these states were coined as spatiotemporal Turing-Hopf pining states. Furthermore, these dynamical states have been found experimentally in resistively coupled nonlinear LC (inductor-capacitor) oscillators \cite{heidemann_fronts_1993}, and in binary fluids in an annular container \cite{kolodner_coexisting_1993}.

\begin{figure*}[!t]
\centering
\includegraphics[scale=1]{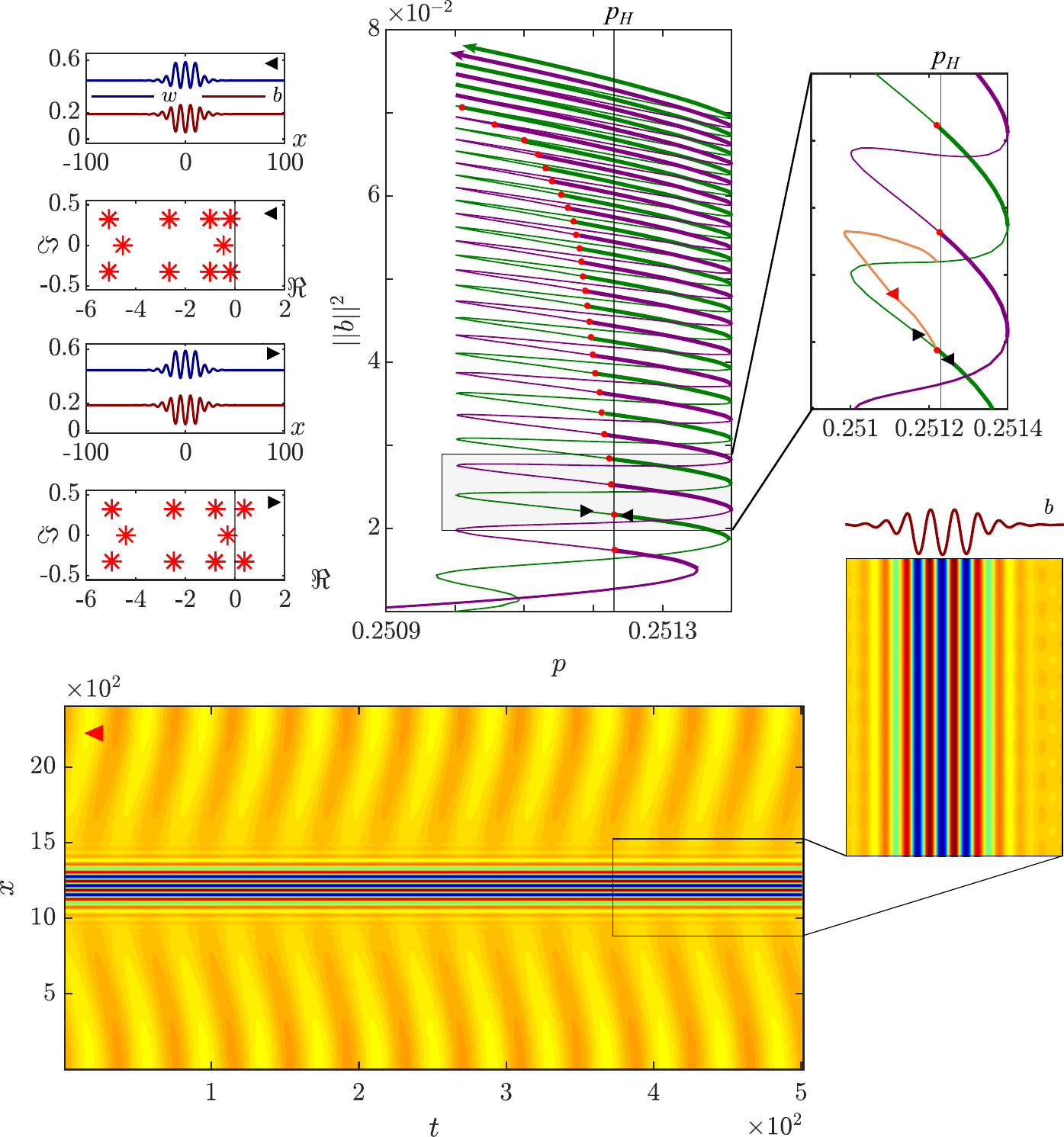}
\caption{Homoclinic bifurcation diagram intersecting the HB (see vertical line at $p_H$) for $n=0.35789$, and $\delta=0.6$. The HB undergone by the different stable branches of the snaking are marked with {\color{red}$\bullet$}. The marker ${\color{red}\blacktriangleright}$ in the close-up view of the snaking corresponds to the TH LS shown below. In the left column three-gap LSs and their eigenspectrum are plotted for values of $p$ just below ($\blacktriangleright$) and above (see $\blacktriangleleft$) HB.   }
\label{SH_Hopf}
\end{figure*}

Here, we present a different scenario where spatiotemporal TH LSs appear due to the interaction between the homoclinic snaking of static gap LSs, associated with a subcritical TI, and the supercritical Hopf bifurcation (HB) undergone by the uniform UV state. Such interaction was absent for the parameter set chosen in previous sections, but can be easily induced by modifying the the diffusion coefficient. An example of that situation is depicted in the $(p,n)$-phase diagram shown in Fig.~\ref{bioep608} for $\delta=0.6$.
For this diffusion value, contrary to the scenario depicted in Fig.~\ref{phase_diagram}, the HB curve (yellow) intersects the homoclinic snaking region associated with the gap LSs (see blue region), destabilizing the previously stable LS branches. The form in which the homoclinic snaking destabilizes is illustrated in Fig.~\ref{SH_Hopf} for $n=0.358$ where we plot the $L_2$-norm
\begin{equation*}
   ||b||^2\equiv L^{-1}\int_{-L/2}^{L/2}b(x)^2dx,
\end{equation*}
as a function of $p$. Here, the HB of the UV state occurs at $p_H=0.25123$ and is marked using the vertical line. 
Proceeding up in the diagram, the previously stable state branches of both $\Gamma_0$ and $\Gamma_\pi$ become unstable through a number of HBs (see red dots in Fig.~\ref{SH_Hopf}). The first two bottom HBs occur for a value very close to $p_H$. However, as the number of pattern rolls in the gap LSs increase (i.e., increasing $||b||^2$), the HBs move to the left, approaching the left folds on the diagram. As a result, the wider the state, the larger its stability range. Eventually, the highest HB collides with a left fold of the snaking diagram and disappears. Above this point, LSs do not undergo any oscillatory instability. 

The left column in Fig.~\ref{SH_Hopf} shows a three-roll biomass gap LS and its eigenspectrum $\sigma$ (i.e., $\mathfrak{R}\equiv{\rm Re}[\sigma]$ versus $\mathfrak{I}\equiv{\rm Im}[\sigma]$) before and after HB (see markers $\blacktriangleleft$ and $\blacktriangleright$ respectively). The modification of the eigenspectrum shows how two complex conjugate eigenvalues crosses the axis $\mathfrak{R}=0$, which is a clear signature of a HB.   

The most right panel in Fig.~\ref{SH_Hopf} shows a close-up view of the center one around $p=p_H$ and the three-roll gap state branch. The orange curve corresponds to the spatiotemporal TH LS emerging from HB, and has been computed through the free distribution continuation software pde2path \cite{uecker_pde2path_2014,uecker_numerical_2021}. An example of such state is depicted at the bottom of Fig.~\ref{SH_Hopf}, where we plot the time evolution of the biomass three-rolls gap state for $b=0.2511$ (see ${\color{red}\blacktriangleright}$ in the diagram). In this state, is the uniform background who oscillates while the localized part of the state remains unchanged. Furthermore, the locations of the interfaces between the Turing and Hopf regions remain constant in time.  This same scenario has been analyzed in other reaction-diffusion systems \cite{de_wit_spatiotemporal_1996,borckmans_localized_1995,tzou_homoclinic_2013}, and is morphologically identical to those previously shown in a supercritical Turing scenario \cite{de_wit_spatiotemporal_1996,borckmans_localized_1995,tzou_homoclinic_2013}.



\section{Discussion and conclusions}\label{sec:5}
In this paper we have performed a detail bifurcation analysis of the 1D simplified Gilad-Meron model for plant ecology in semi-arid landscapes. This model has 7 control parameters, what makes its full analysis quite cumbersome. In order to reduce this complexity, we have fixed 5 of them, and varied the parameters related with the precipitations $p$ and evaporation $n$ rates, which are essential to understand the vegetation dynamics. Despite the large amount of work dedicated to this model and its variations, a systematic study of the bifurcation structure and stability of their LSs, as well as, their taxonomical  organization in the parameter space was lacking.  

Localized states normally form due to the locking of fronts connecting two different, but coexisting, states, which could be uniform or not. 
We have analyze the formation of LSs in two main configurations: 
\begin{itemize}
  \item The UV uniform state coexists with a subcritical Turing pattern, and the locking of fronts yield LSs consisting in a slug of the spatially periodic pattern, containing a given number of pattern rolls, embedded in UV. Here, LSs consist in localized vegetation gaps. These states undergo standard homoclinic snaking like the one studied in Sec.~\ref{sec:3B}.

    \item Two coexisting uniform states such as BS and UV coexist. Here, the locking of fronts connecting such states lead to the formation of LSs consisting in a plateau of UV embedded in the BS state. These spots or biomass are organized in a collapsed homoclinic snaking bifurcation structure. This bifurcation scenario appears for high values of $n$ [see phase diagram in Fig.~\ref{phase_diagram}], and has been analyzed in Sec.~\ref{sec:3D}.


\end{itemize}
While homoclinic snaking has been studied in the simplified GM model for a different regime of parameters where the biomass LSs are spots\cite{zelnik_desertification_2017}, to our knowledge, this is the first time that collapsed snaking is reported. 
The transition between these scenarios occurs in a region where tristability between the BS, UV and the subcritical Turing pattern states exists. We have found that this multi-coexistence leads to a plethora of hybrid states and to the appearance of different type of isolas that we have studied in Sec.~\ref{sec:3C}.
This tristable scenario was first pointed out by Zelnik {\it et al.} in the context of the Klausmeier–Gray–Scott
model \cite{zelnik_implications_2018}, and similar configurations have been found in other reaction-diffusion models such as the Gierer-Minhardt model \cite{al_saadi_unified_2021}, and in the Thomas model for enzyme reactors \cite{ALSAADI2022}. 

We have found that the reorganization of the bifurcation curves is leaded by necking bifurcations which reconnect isolas belonging to a different LS class. This is a generic transition in pattern forming systems with tristability general\cite{parra-rivas_organization_2022}.

The UV state undergoes a Hopf bifurcation, which, in the absence of diffusion, develops periodic oscillations. When diffusion is considered TI comes into play, leading to complex dynamical scenarios where the Turing and Hopf modes interact \cite{de_wit_spatiotemporal_1996}. The last part of our work (see Sec.~\ref{sec:4}) has focused on the analysis of how the stability and dynamics of gap biomass LSs modify when entering the TH instability region.  We have found that the uniform HB destabilizes the homoclinic snaking states, leading to complex spatiotemporal LSs consisting in a localized pattern of a few peaks embedded in uniform oscillations of the UV state (see Fig.~\ref{SH_Hopf}). For LSs with a very few peaks such instability develops at the same $p$ value that the uniform one. Increasing the number of peaks, however, the HB of each branch moves toward $p<p_H$, stabilizing sequentially the localized vegetation gaps of different extensions. This sequential stabilization has been also found in other systems not involving uniform HBs \cite{parra-rivas_parametric_2020}.

In summary, the results presented here offer a clear classification of the different type of states appearing on this model, and potentially in the ecosystems described by it. This was a necessary work in order to fully understand the transition between bifurcation scenarios and LSs which has ramifications non only in plant ecology, but in many other fields of nonlinear science. 

\section*{The conflict of interest statement }
This work does not have any conflicts of interest.
\section*{Acknowledgements}
P. P. -R acknowledges support from the European Union’s Horizon 2020 research and innovation programme
under the Marie Sklodowska-Curie grant agreement no. 101023717.

\bibliography{mybibfile}

\begin{thebibliography}{65}%
\makeatletter
\providecommand \@ifxundefined [1]{%
 \@ifx{#1\undefined}
}%
\providecommand \@ifnum [1]{%
 \ifnum #1\expandafter \@firstoftwo
 \else \expandafter \@secondoftwo
 \fi
}%
\providecommand \@ifx [1]{%
 \ifx #1\expandafter \@firstoftwo
 \else \expandafter \@secondoftwo
 \fi
}%
\providecommand \natexlab [1]{#1}%
\providecommand \enquote  [1]{``#1''}%
\providecommand \bibnamefont  [1]{#1}%
\providecommand \bibfnamefont [1]{#1}%
\providecommand \citenamefont [1]{#1}%
\providecommand \href@noop [0]{\@secondoftwo}%
\providecommand \href [0]{\begingroup \@sanitize@url \@href}%
\providecommand \@href[1]{\@@startlink{#1}\@@href}%
\providecommand \@@href[1]{\endgroup#1\@@endlink}%
\providecommand \@sanitize@url [0]{\catcode `\\12\catcode `\$12\catcode
  `\&12\catcode `\#12\catcode `\^12\catcode `\_12\catcode `\%12\relax}%
\providecommand \@@startlink[1]{}%
\providecommand \@@endlink[0]{}%
\providecommand \url  [0]{\begingroup\@sanitize@url \@url }%
\providecommand \@url [1]{\endgroup\@href {#1}{\urlprefix }}%
\providecommand \urlprefix  [0]{URL }%
\providecommand \Eprint [0]{\href }%
\providecommand \doibase [0]{https://doi.org/}%
\providecommand \selectlanguage [0]{\@gobble}%
\providecommand \bibinfo  [0]{\@secondoftwo}%
\providecommand \bibfield  [0]{\@secondoftwo}%
\providecommand \translation [1]{[#1]}%
\providecommand \BibitemOpen [0]{}%
\providecommand \bibitemStop [0]{}%
\providecommand \bibitemNoStop [0]{.\EOS\space}%
\providecommand \EOS [0]{\spacefactor3000\relax}%
\providecommand \BibitemShut  [1]{\csname bibitem#1\endcsname}%
\let\auto@bib@innerbib\@empty
\bibitem [{\citenamefont {Cross}\ and\ \citenamefont
  {Hohenberg}(1993)}]{cross_pattern_1993}%
  \BibitemOpen
  \bibfield  {author} {\bibinfo {author} {\bibfnamefont {M.~C.}\ \bibnamefont
  {Cross}}\ and\ \bibinfo {author} {\bibfnamefont {P.~C.}\ \bibnamefont
  {Hohenberg}},\ }\bibfield  {title} {\enquote {\bibinfo {title} {Pattern
  formation outside of equilibrium},}\ }\href
  {https://doi.org/10.1103/RevModPhys.65.851} {\bibfield  {journal} {\bibinfo
  {journal} {Reviews of Modern Physics}\ }\textbf {\bibinfo {volume} {65}},\
  \bibinfo {pages} {851--1112} (\bibinfo {year} {1993})}\BibitemShut {NoStop}%
\bibitem [{\citenamefont {Cross}\ and\ \citenamefont
  {Greenside}(2009)}]{cross_pattern_2009}%
  \BibitemOpen
  \bibfield  {author} {\bibinfo {author} {\bibfnamefont {M.}~\bibnamefont
  {Cross}}\ and\ \bibinfo {author} {\bibfnamefont {H.}~\bibnamefont
  {Greenside}},\ }\href {https://doi.org/10.1017/CBO9780511627200} {\emph
  {\bibinfo {title} {Pattern {Formation} and {Dynamics} in {Nonequilibrium}
  {Systems}}}}\ (\bibinfo  {publisher} {Cambridge University Press},\ \bibinfo
  {address} {Cambridge},\ \bibinfo {year} {2009})\BibitemShut {NoStop}%
\bibitem [{\citenamefont {Akhmediev}\ and\ \citenamefont
  {Ankiewicz}(2008)}]{akhmediev_dissipative_2008}%
  \BibitemOpen
  \bibinfo {editor} {\bibfnamefont {N.}~\bibnamefont {Akhmediev}}\ and\
  \bibinfo {editor} {\bibfnamefont {A.}~\bibnamefont {Ankiewicz}},\ eds.,\
  \href {https://doi.org/10.1007/978-3-540-78217-9} {\emph {\bibinfo {title}
  {Dissipative {Solitons}: {From} {Optics} to {Biology} and {Medicine}}}},\
  Lecture {Notes} in {Physics}\ (\bibinfo  {publisher} {Springer-Verlag},\
  \bibinfo {address} {Berlin Heidelberg},\ \bibinfo {year} {2008})\BibitemShut
  {NoStop}%
\bibitem [{\citenamefont {Descalzi}\ \emph {et~al.}(2011)\citenamefont
  {Descalzi}, \citenamefont {Clerc}, \citenamefont {Residori},\ and\
  \citenamefont {Assanto}}]{descalzi_localized_2011}%
  \BibitemOpen
  \bibinfo {editor} {\bibfnamefont {O.}~\bibnamefont {Descalzi}}, \bibinfo
  {editor} {\bibfnamefont {M.~G.}\ \bibnamefont {Clerc}}, \bibinfo {editor}
  {\bibfnamefont {S.}~\bibnamefont {Residori}},\ and\ \bibinfo {editor}
  {\bibfnamefont {G.}~\bibnamefont {Assanto}},\ eds.,\ \href
  {https://doi.org/10.1007/978-3-642-16549-8} {\emph {\bibinfo {title}
  {Localized {States} in {Physics}: {Solitons} and {Patterns}}}}\ (\bibinfo
  {publisher} {Springer-Verlag},\ \bibinfo {address} {Berlin Heidelberg},\
  \bibinfo {year} {2011})\BibitemShut {NoStop}%
\bibitem [{\citenamefont {Knobloch}(2015)}]{knobloch_spatial_2015}%
  \BibitemOpen
  \bibfield  {author} {\bibinfo {author} {\bibfnamefont {E.}~\bibnamefont
  {Knobloch}},\ }\bibfield  {title} {\enquote {\bibinfo {title} {Spatial
  {Localization} in {Dissipative} {Systems}},}\ }\href
  {https://doi.org/10.1146/annurev-conmatphys-031214-014514} {\bibfield
  {journal} {\bibinfo  {journal} {Annual Review of Condensed Matter Physics}\
  }\textbf {\bibinfo {volume} {6}},\ \bibinfo {pages} {325--359} (\bibinfo
  {year} {2015})}\BibitemShut {NoStop}%
\bibitem [{\citenamefont {Al~Saadi}\ \emph {et~al.}(2021)\citenamefont
  {Al~Saadi}, \citenamefont {Champneys}, \citenamefont {Worthy},\ and\
  \citenamefont {Msmali}}]{Fahadpray}%
  \BibitemOpen
  \bibfield  {author} {\bibinfo {author} {\bibfnamefont {F.}~\bibnamefont
  {Al~Saadi}}, \bibinfo {author} {\bibfnamefont {A.}~\bibnamefont {Champneys}},
  \bibinfo {author} {\bibfnamefont {A.}~\bibnamefont {Worthy}},\ and\ \bibinfo
  {author} {\bibfnamefont {A.}~\bibnamefont {Msmali}},\ }\bibfield  {title}
  {\enquote {\bibinfo {title} {{Stationary and oscillatory localized patterns
  in ratio-dependent predator–prey systems}},}\ }\href@noop {} {\bibfield
  {journal} {\bibinfo  {journal} {IMA Journal of Applied Mathematics}\ }\textbf
  {\bibinfo {volume} {86}},\ \bibinfo {pages} {808--827} (\bibinfo {year}
  {2021})}\BibitemShut {NoStop}%
\bibitem [{\citenamefont {Coullet}, \citenamefont {Elphick},\ and\
  \citenamefont {Repaux}(1987)}]{coullet_nature_1987}%
  \BibitemOpen
  \bibfield  {author} {\bibinfo {author} {\bibfnamefont {P.}~\bibnamefont
  {Coullet}}, \bibinfo {author} {\bibfnamefont {C.}~\bibnamefont {Elphick}},\
  and\ \bibinfo {author} {\bibfnamefont {D.}~\bibnamefont {Repaux}},\
  }\bibfield  {title} {\enquote {\bibinfo {title} {Nature of spatial chaos},}\
  }\href@noop {} {\bibfield  {journal} {\bibinfo  {journal} {PHYSICAL REVIEW
  LETTERS}\ }\textbf {\bibinfo {volume} {58}},\ \bibinfo {pages} {4} (\bibinfo
  {year} {1987})}\BibitemShut {NoStop}%
\bibitem [{\citenamefont {Thual}\ and\ \citenamefont
  {Fauve}(1988)}]{thual_localized_1988}%
  \BibitemOpen
  \bibfield  {author} {\bibinfo {author} {\bibfnamefont {O.}~\bibnamefont
  {Thual}}\ and\ \bibinfo {author} {\bibfnamefont {S.}~\bibnamefont {Fauve}},\
  }\bibfield  {title} {\enquote {\bibinfo {title} {Localized structures
  generated by subcritical instabilities},}\ }\href
  {https://doi.org/10.1051/jphys:0198800490110182900} {\bibfield  {journal}
  {\bibinfo  {journal} {Journal de Physique}\ }\textbf {\bibinfo {volume}
  {49}},\ \bibinfo {pages} {1829--1833} (\bibinfo {year} {1988})},\ \bibinfo
  {note} {publisher: Société Française de Physique}\BibitemShut {NoStop}%
\bibitem [{\citenamefont {Coullet}(2002)}]{coullet_localized_2002}%
  \BibitemOpen
  \bibfield  {author} {\bibinfo {author} {\bibfnamefont {P.}~\bibnamefont
  {Coullet}},\ }\bibfield  {title} {\enquote {\bibinfo {title} {Localized
  patterns and fronts in nonequilibrium systems},}\ }\href
  {https://doi.org/10.1142/S021812740200614X} {\bibfield  {journal} {\bibinfo
  {journal} {International Journal of Bifurcation and Chaos}\ }\textbf
  {\bibinfo {volume} {12}},\ \bibinfo {pages} {2445--2457} (\bibinfo {year}
  {2002})}\BibitemShut {NoStop}%
\bibitem [{\citenamefont {Macfadyen}(1950)}]{macfadyen_vegetation_1950}%
  \BibitemOpen
  \bibfield  {author} {\bibinfo {author} {\bibfnamefont {W.~A.}\ \bibnamefont
  {Macfadyen}},\ }\bibfield  {title} {\enquote {\bibinfo {title} {Vegetation
  patterns in the semi-desert plains of {British} {Somaliland}},}\ }\href
  {https://doi.org/10.2307/1789384} {\bibfield  {journal} {\bibinfo  {journal}
  {The Geographical Journal}\ }\textbf {\bibinfo {volume} {116}},\ \bibinfo
  {pages} {199--211} (\bibinfo {year} {1950})}\BibitemShut {NoStop}%
\bibitem [{\citenamefont {Becker}\ and\ \citenamefont
  {Getzin}(2000)}]{becker_fairy_2000}%
  \BibitemOpen
  \bibfield  {author} {\bibinfo {author} {\bibfnamefont {T.}~\bibnamefont
  {Becker}}\ and\ \bibinfo {author} {\bibfnamefont {S.}~\bibnamefont
  {Getzin}},\ }\bibfield  {title} {\enquote {\bibinfo {title} {The fairy
  circles of {Kaokoland} ({North}-{West} {Namibia}) origin, distribution, and
  characteristics},}\ }\href {https://doi.org/10.1078/1439-1791-00021}
  {\bibfield  {journal} {\bibinfo  {journal} {Basic and Applied Ecology}\
  }\textbf {\bibinfo {volume} {1}},\ \bibinfo {pages} {149--159} (\bibinfo
  {year} {2000})}\BibitemShut {NoStop}%
\bibitem [{\citenamefont {van Rooyen}\ \emph {et~al.}(2004)\citenamefont {van
  Rooyen}, \citenamefont {Theron}, \citenamefont {van Rooyen}, \citenamefont
  {Jankowitz},\ and\ \citenamefont {Matthews}}]{van_rooyen_mysterious_2004}%
  \BibitemOpen
  \bibfield  {author} {\bibinfo {author} {\bibfnamefont {M.~W.}\ \bibnamefont
  {van Rooyen}}, \bibinfo {author} {\bibfnamefont {G.~K.}\ \bibnamefont
  {Theron}}, \bibinfo {author} {\bibfnamefont {N.}~\bibnamefont {van Rooyen}},
  \bibinfo {author} {\bibfnamefont {W.~J.}\ \bibnamefont {Jankowitz}},\ and\
  \bibinfo {author} {\bibfnamefont {W.~S.}\ \bibnamefont {Matthews}},\
  }\bibfield  {title} {\enquote {\bibinfo {title} {Mysterious circles in the
  {Namib} {Desert}: Review of hypotheses on their origin},}\ }\href
  {https://doi.org/10.1016/S0140-1963(03)00111-3} {\bibfield  {journal}
  {\bibinfo  {journal} {Journal of Arid Environments}\ }\textbf {\bibinfo
  {volume} {57}},\ \bibinfo {pages} {467--485} (\bibinfo {year}
  {2004})}\BibitemShut {NoStop}%
\bibitem [{\citenamefont {Meron}, \citenamefont {Yizhaq},\ and\ \citenamefont
  {Gilad}(2007)}]{meron_localized_2007}%
  \BibitemOpen
  \bibfield  {author} {\bibinfo {author} {\bibfnamefont {E.}~\bibnamefont
  {Meron}}, \bibinfo {author} {\bibfnamefont {H.}~\bibnamefont {Yizhaq}},\ and\
  \bibinfo {author} {\bibfnamefont {E.}~\bibnamefont {Gilad}},\ }\bibfield
  {title} {\enquote {\bibinfo {title} {Localized structures in dryland
  vegetation: {Forms} and functions},}\ }\href
  {https://doi.org/10.1063/1.2767246} {\bibfield  {journal} {\bibinfo
  {journal} {Chaos: An Interdisciplinary Journal of Nonlinear Science}\
  }\textbf {\bibinfo {volume} {17}},\ \bibinfo {pages} {037109} (\bibinfo
  {year} {2007})}\BibitemShut {NoStop}%
\bibitem [{\citenamefont {Deblauwe}\ \emph {et~al.}(2011)\citenamefont
  {Deblauwe}, \citenamefont {Couteron}, \citenamefont {Lejeune}, \citenamefont
  {Bogaert},\ and\ \citenamefont {Barbier}}]{VegBxl2011Deblauwe}%
  \BibitemOpen
  \bibfield  {author} {\bibinfo {author} {\bibfnamefont {V.}~\bibnamefont
  {Deblauwe}}, \bibinfo {author} {\bibfnamefont {P.}~\bibnamefont {Couteron}},
  \bibinfo {author} {\bibfnamefont {O.}~\bibnamefont {Lejeune}}, \bibinfo
  {author} {\bibfnamefont {J.}~\bibnamefont {Bogaert}},\ and\ \bibinfo {author}
  {\bibfnamefont {N.}~\bibnamefont {Barbier}},\ }\bibfield  {title} {\enquote
  {\bibinfo {title} {Environmental modulation of self-organized periodic
  vegetation patterns in {S}udan},}\ }\href
  {https://doi.org/10.1111/j.1600-0587.2010.06694.x} {\bibfield  {journal}
  {\bibinfo  {journal} {Ecography}\ }\textbf {\bibinfo {volume} {34}},\
  \bibinfo {pages} {990--1001} (\bibinfo {year} {2011})}\BibitemShut {NoStop}%
\bibitem [{\citenamefont {Meron}(2012)}]{meron_pattern-formation_2012}%
  \BibitemOpen
  \bibfield  {author} {\bibinfo {author} {\bibfnamefont {E.}~\bibnamefont
  {Meron}},\ }\bibfield  {title} {\enquote {\bibinfo {title} {Pattern-formation
  approach to modelling spatially extended ecosystems},}\ }\href
  {https://doi.org/10.1016/j.ecolmodel.2011.05.035} {\bibfield  {journal}
  {\bibinfo  {journal} {Ecological Modelling}\ }\bibinfo {series} {Modelling
  clonal plant growth: {From} {Ecological} concepts to {Mathematics}},\ \textbf
  {\bibinfo {volume} {234}},\ \bibinfo {pages} {70--82} (\bibinfo {year}
  {2012})}\BibitemShut {NoStop}%
\bibitem [{\citenamefont {Tschinkel}(2015)}]{tschinkel_experiments_2015}%
  \BibitemOpen
  \bibfield  {author} {\bibinfo {author} {\bibfnamefont {W.~R.}\ \bibnamefont
  {Tschinkel}},\ }\bibfield  {title} {\enquote {\bibinfo {title} {Experiments
  testing the causes of {Namibian} fairy circles},}\ }\href
  {https://doi.org/10.1371/journal.pone.0140099} {\bibfield  {journal}
  {\bibinfo  {journal} {PLOS ONE}\ }\textbf {\bibinfo {volume} {10}},\ \bibinfo
  {pages} {e0140099} (\bibinfo {year} {2015})}\BibitemShut {NoStop}%
\bibitem [{\citenamefont {Getzin}\ \emph {et~al.}(2016)\citenamefont {Getzin},
  \citenamefont {Yizhaq}, \citenamefont {Bell}, \citenamefont {Erickson},
  \citenamefont {Postle}, \citenamefont {Katra}, \citenamefont {Tzuk},
  \citenamefont {Zelnik}, \citenamefont {Wiegand}, \citenamefont {Wiegand},\
  and\ \citenamefont {Meron}}]{getzin_discovery_2016}%
  \BibitemOpen
  \bibfield  {author} {\bibinfo {author} {\bibfnamefont {S.}~\bibnamefont
  {Getzin}}, \bibinfo {author} {\bibfnamefont {H.}~\bibnamefont {Yizhaq}},
  \bibinfo {author} {\bibfnamefont {B.}~\bibnamefont {Bell}}, \bibinfo {author}
  {\bibfnamefont {T.~E.}\ \bibnamefont {Erickson}}, \bibinfo {author}
  {\bibfnamefont {A.~C.}\ \bibnamefont {Postle}}, \bibinfo {author}
  {\bibfnamefont {I.}~\bibnamefont {Katra}}, \bibinfo {author} {\bibfnamefont
  {O.}~\bibnamefont {Tzuk}}, \bibinfo {author} {\bibfnamefont {Y.~R.}\
  \bibnamefont {Zelnik}}, \bibinfo {author} {\bibfnamefont {K.}~\bibnamefont
  {Wiegand}}, \bibinfo {author} {\bibfnamefont {T.}~\bibnamefont {Wiegand}},\
  and\ \bibinfo {author} {\bibfnamefont {E.}~\bibnamefont {Meron}},\ }\bibfield
   {title} {\enquote {\bibinfo {title} {Discovery of fairy circles in
  {Australia} supports self-organization theory},}\ }\href
  {https://doi.org/10.1073/pnas.1522130113} {\bibfield  {journal} {\bibinfo
  {journal} {Proceedings of the National Academy of Sciences}\ }\textbf
  {\bibinfo {volume} {113}},\ \bibinfo {pages} {3551--3556} (\bibinfo {year}
  {2016})}\BibitemShut {NoStop}%
\bibitem [{\citenamefont {Ruiz-Reyn\'es}\ \emph {et~al.}(2017)\citenamefont
  {Ruiz-Reyn\'es}, \citenamefont {Gomila}, \citenamefont {Sintes},
  \citenamefont {Hern\'andez-Garc\'ia}, \citenamefont {Marbà},\ and\
  \citenamefont {Duarte}}]{ruiz-reynes_fairy_2017}%
  \BibitemOpen
  \bibfield  {author} {\bibinfo {author} {\bibfnamefont {D.}~\bibnamefont
  {Ruiz-Reyn\'es}}, \bibinfo {author} {\bibfnamefont {D.}~\bibnamefont
  {Gomila}}, \bibinfo {author} {\bibfnamefont {T.}~\bibnamefont {Sintes}},
  \bibinfo {author} {\bibfnamefont {E.}~\bibnamefont {Hern\'andez-Garc\'ia}},
  \bibinfo {author} {\bibfnamefont {N.}~\bibnamefont {Marbà}},\ and\ \bibinfo
  {author} {\bibfnamefont {C.~M.}\ \bibnamefont {Duarte}},\ }\bibfield  {title}
  {\enquote {\bibinfo {title} {Fairy circle landscapes under the sea},}\ }\href
  {https://doi.org/10.1126/sciadv.1603262} {\bibfield  {journal} {\bibinfo
  {journal} {Science Advances}\ }\textbf {\bibinfo {volume} {3}},\ \bibinfo
  {pages} {e1603262} (\bibinfo {year} {2017})}\BibitemShut {NoStop}%
\bibitem [{\citenamefont {Lejeune}, \citenamefont {Tlidi},\ and\ \citenamefont
  {Couteron}(2002)}]{lejeune_localized_2002}%
  \BibitemOpen
  \bibfield  {author} {\bibinfo {author} {\bibfnamefont {O.}~\bibnamefont
  {Lejeune}}, \bibinfo {author} {\bibfnamefont {M.}~\bibnamefont {Tlidi}},\
  and\ \bibinfo {author} {\bibfnamefont {P.}~\bibnamefont {Couteron}},\
  }\bibfield  {title} {\enquote {\bibinfo {title} {Localized vegetation
  patches: {A} self-organized response to resource scarcity},}\ }\href
  {https://doi.org/10.1103/PhysRevE.66.010901} {\bibfield  {journal} {\bibinfo
  {journal} {Physical Review E}\ }\textbf {\bibinfo {volume} {66}},\ \bibinfo
  {pages} {010901} (\bibinfo {year} {2002})}\BibitemShut {NoStop}%
\bibitem [{\citenamefont {Escaff}\ \emph {et~al.}(2015)\citenamefont {Escaff},
  \citenamefont {Fern\'andez-Oto}, \citenamefont {Clerc},\ and\ \citenamefont
  {Tlidi}}]{escaff_localized_2015}%
  \BibitemOpen
  \bibfield  {author} {\bibinfo {author} {\bibfnamefont {D.}~\bibnamefont
  {Escaff}}, \bibinfo {author} {\bibfnamefont {C.}~\bibnamefont
  {Fern\'andez-Oto}}, \bibinfo {author} {\bibfnamefont {M.~G.}\ \bibnamefont
  {Clerc}},\ and\ \bibinfo {author} {\bibfnamefont {M.}~\bibnamefont {Tlidi}},\
  }\bibfield  {title} {\enquote {\bibinfo {title} {Localized vegetation
  patterns, fairy circles, and localized patches in arid landscapes},}\ }\href
  {https://doi.org/10.1103/PhysRevE.91.022924} {\bibfield  {journal} {\bibinfo
  {journal} {Physical Review E}\ }\textbf {\bibinfo {volume} {91}},\ \bibinfo
  {pages} {022924} (\bibinfo {year} {2015})}\BibitemShut {NoStop}%
\bibitem [{\citenamefont {Zelnik}\ \emph {et~al.}(2013)\citenamefont {Zelnik},
  \citenamefont {Kinast}, \citenamefont {Yizhaq}, \citenamefont {Bel},\ and\
  \citenamefont {Meron}}]{zelnik_yuval_r._regime_2013}%
  \BibitemOpen
  \bibfield  {author} {\bibinfo {author} {\bibfnamefont {Y.~R.}\ \bibnamefont
  {Zelnik}}, \bibinfo {author} {\bibfnamefont {S.}~\bibnamefont {Kinast}},
  \bibinfo {author} {\bibfnamefont {H.}~\bibnamefont {Yizhaq}}, \bibinfo
  {author} {\bibfnamefont {G.}~\bibnamefont {Bel}},\ and\ \bibinfo {author}
  {\bibfnamefont {E.}~\bibnamefont {Meron}},\ }\bibfield  {title} {\enquote
  {\bibinfo {title} {Regime shifts in models of dryland vegetation},}\ }\href
  {https://doi.org/10.1098/rsta.2012.0358} {\bibfield  {journal} {\bibinfo
  {journal} {Philosophical Transactions of the Royal Society A: Mathematical,
  Physical and Engineering Sciences}\ }\textbf {\bibinfo {volume} {371}},\
  \bibinfo {pages} {20120358} (\bibinfo {year} {2013})}\BibitemShut {NoStop}%
\bibitem [{\citenamefont {Tlidi}, \citenamefont {Lefever},\ and\ \citenamefont
  {Vladimirov}(2008)}]{tlidi_vegetation_2008}%
  \BibitemOpen
  \bibfield  {author} {\bibinfo {author} {\bibfnamefont {M.}~\bibnamefont
  {Tlidi}}, \bibinfo {author} {\bibfnamefont {R.}~\bibnamefont {Lefever}},\
  and\ \bibinfo {author} {\bibfnamefont {A.}~\bibnamefont {Vladimirov}},\
  }\bibfield  {title} {\enquote {\bibinfo {title} {On vegetation clustering,
  localized bare soil spots and fairy circles},}\ }in\ \href
  {https://doi.org/10.1007/978-3-540-78217-9_15} {\emph {\bibinfo {booktitle}
  {Dissipative {Solitons}: {From} {Optics} to {Biology} and {Medicine}}}},\
  \bibinfo {series and number} {Lecture {Notes} in {Physics}}\ (\bibinfo
  {publisher} {Springer Berlin Heidelberg},\ \bibinfo {address} {Berlin,
  Heidelberg},\ \bibinfo {year} {2008})\ pp.\ \bibinfo {pages}
  {1--22}\BibitemShut {NoStop}%
\bibitem [{\citenamefont {Fern\'andez-Oto}\ \emph {et~al.}(2014)\citenamefont
  {Fern\'andez-Oto}, \citenamefont {Tlidi}, \citenamefont {Escaff},\ and\
  \citenamefont {Clerc}}]{Veg_FC_2014_Fernandez-Oto}%
  \BibitemOpen
  \bibfield  {author} {\bibinfo {author} {\bibfnamefont {C.}~\bibnamefont
  {Fern\'andez-Oto}}, \bibinfo {author} {\bibfnamefont {M.}~\bibnamefont
  {Tlidi}}, \bibinfo {author} {\bibfnamefont {D.}~\bibnamefont {Escaff}},\ and\
  \bibinfo {author} {\bibfnamefont {M.~G.}\ \bibnamefont {Clerc}},\ }\bibfield
  {title} {\enquote {\bibinfo {title} {Strong interaction between plants
  induces circular barren patches: {F}airy circles},}\ }\href
  {https://doi.org/10.1098/rsta.2014.0009} {\bibfield  {journal} {\bibinfo
  {journal} {Philosophical Transactions of the Royal Society of London A:
  Mathematical, Physical and Engineering Sciences}\ }\textbf {\bibinfo {volume}
  {372}},\ \bibinfo {pages} {20140009} (\bibinfo {year} {2014})},\ \Eprint
  {https://arxiv.org/abs/http://rsta.royalsocietypublishing.org/content/372/2027/20140009.full.pdf}
  {http://rsta.royalsocietypublishing.org/content/372/2027/20140009.full.pdf}
  \BibitemShut {NoStop}%
\bibitem [{\citenamefont {Zelnik}, \citenamefont {Meron},\ and\ \citenamefont
  {Bel}(2016)}]{zelnik_localized_2016}%
  \BibitemOpen
  \bibfield  {author} {\bibinfo {author} {\bibfnamefont {Y.~R.}\ \bibnamefont
  {Zelnik}}, \bibinfo {author} {\bibfnamefont {E.}~\bibnamefont {Meron}},\ and\
  \bibinfo {author} {\bibfnamefont {G.}~\bibnamefont {Bel}},\ }\bibfield
  {title} {\enquote {\bibinfo {title} {Localized states qualitatively change
  the response of ecosystems to varying conditions and local disturbances},}\
  }\href {https://doi.org/10.1016/j.ecocom.2015.11.004} {\bibfield  {journal}
  {\bibinfo  {journal} {Ecological Complexity}\ }\textbf {\bibinfo {volume}
  {25}},\ \bibinfo {pages} {26--34} (\bibinfo {year} {2016})}\BibitemShut
  {NoStop}%
\bibitem [{\citenamefont {Sheffer}\ \emph {et~al.}(2007)\citenamefont
  {Sheffer}, \citenamefont {Yizhaq}, \citenamefont {Gilad}, \citenamefont
  {Shachak},\ and\ \citenamefont {Meron}}]{VegIsrael2007Sheffer2007}%
  \BibitemOpen
  \bibfield  {author} {\bibinfo {author} {\bibfnamefont {E.}~\bibnamefont
  {Sheffer}}, \bibinfo {author} {\bibfnamefont {H.}~\bibnamefont {Yizhaq}},
  \bibinfo {author} {\bibfnamefont {E.}~\bibnamefont {Gilad}}, \bibinfo
  {author} {\bibfnamefont {M.}~\bibnamefont {Shachak}},\ and\ \bibinfo {author}
  {\bibfnamefont {E.}~\bibnamefont {Meron}},\ }\bibfield  {title} {\enquote
  {\bibinfo {title} {Why do plants in resource-deprived environments form
  rings?}}\ }\href
  {https://doi.org/http://dx.doi.org/10.1016/j.ecocom.2007.06.008} {\bibfield
  {journal} {\bibinfo  {journal} {Ecological Complexity}\ }\textbf {\bibinfo
  {volume} {4}},\ \bibinfo {pages} {192 -- 200} (\bibinfo {year}
  {2007})}\BibitemShut {NoStop}%
\bibitem [{\citenamefont {Sheffer}\ \emph {et~al.}(2011)\citenamefont
  {Sheffer}, \citenamefont {Yizhaq}, \citenamefont {Shachak},\ and\
  \citenamefont {Meron}}]{VegIsrael2011Sheffer}%
  \BibitemOpen
  \bibfield  {author} {\bibinfo {author} {\bibfnamefont {E.}~\bibnamefont
  {Sheffer}}, \bibinfo {author} {\bibfnamefont {H.}~\bibnamefont {Yizhaq}},
  \bibinfo {author} {\bibfnamefont {M.}~\bibnamefont {Shachak}},\ and\ \bibinfo
  {author} {\bibfnamefont {E.}~\bibnamefont {Meron}},\ }\bibfield  {title}
  {\enquote {\bibinfo {title} {Mechanisms of vegetation-ring formation in
  water-limited systems},}\ }\href
  {https://doi.org/http://dx.doi.org/10.1016/j.jtbi.2010.12.028} {\bibfield
  {journal} {\bibinfo  {journal} {Journal of Theoretical Biology}\ }\textbf
  {\bibinfo {volume} {273}},\ \bibinfo {pages} {138 -- 146} (\bibinfo {year}
  {2011})}\BibitemShut {NoStop}%
\bibitem [{\citenamefont {Yizhaq}\ \emph {et~al.}()\citenamefont {Yizhaq},
  \citenamefont {Stavi}, \citenamefont {Swet}, \citenamefont {Zaady},\ and\
  \citenamefont {Katra}}]{VegIsrael2019YizhaqRings}%
  \BibitemOpen
  \bibfield  {author} {\bibinfo {author} {\bibfnamefont {H.}~\bibnamefont
  {Yizhaq}}, \bibinfo {author} {\bibfnamefont {I.}~\bibnamefont {Stavi}},
  \bibinfo {author} {\bibfnamefont {N.}~\bibnamefont {Swet}}, \bibinfo {author}
  {\bibfnamefont {E.}~\bibnamefont {Zaady}},\ and\ \bibinfo {author}
  {\bibfnamefont {I.}~\bibnamefont {Katra}},\ }\bibfield  {title} {\enquote
  {\bibinfo {title} {Vegetation ring formation by water overland flow in
  water-limited environments: Field measurements and mathematical modeling},}\
  }\href {https://doi.org/10.1002/eco.2135} {\bibfield  {journal} {\bibinfo
  {journal} {Ecohydrology}\ }\textbf {\bibinfo {volume} {0}},\ \bibinfo {pages}
  {e2135}},\ \bibinfo {note} {e2135 ECO-19-0001.R1},\ \Eprint
  {https://arxiv.org/abs/https://onlinelibrary.wiley.com/doi/pdf/10.1002/eco.2135}
  {https://onlinelibrary.wiley.com/doi/pdf/10.1002/eco.2135} \BibitemShut
  {NoStop}%
\bibitem [{\citenamefont {Zelnik}\ \emph {et~al.}(2017)\citenamefont {Zelnik},
  \citenamefont {Uecker}, \citenamefont {Feudel},\ and\ \citenamefont
  {Meron}}]{zelnik_desertification_2017}%
  \BibitemOpen
  \bibfield  {author} {\bibinfo {author} {\bibfnamefont {Y.~R.}\ \bibnamefont
  {Zelnik}}, \bibinfo {author} {\bibfnamefont {H.}~\bibnamefont {Uecker}},
  \bibinfo {author} {\bibfnamefont {U.}~\bibnamefont {Feudel}},\ and\ \bibinfo
  {author} {\bibfnamefont {E.}~\bibnamefont {Meron}},\ }\bibfield  {title}
  {\enquote {\bibinfo {title} {Desertification by front propagation?}}\ }\href
  {https://doi.org/10.1016/j.jtbi.2017.01.029} {\bibfield  {journal} {\bibinfo
  {journal} {Journal of Theoretical Biology}\ }\textbf {\bibinfo {volume}
  {418}},\ \bibinfo {pages} {27--35} (\bibinfo {year} {2017})}\BibitemShut
  {NoStop}%
\bibitem [{\citenamefont {Scheffer}\ \emph {et~al.}(2001)\citenamefont
  {Scheffer}, \citenamefont {Carpenter}, \citenamefont {Foley}, \citenamefont
  {Folke},\ and\ \citenamefont {Walker}}]{VegOtros2012Scheffer}%
  \BibitemOpen
  \bibfield  {author} {\bibinfo {author} {\bibfnamefont {M.}~\bibnamefont
  {Scheffer}}, \bibinfo {author} {\bibfnamefont {S.}~\bibnamefont {Carpenter}},
  \bibinfo {author} {\bibfnamefont {J.~A.}\ \bibnamefont {Foley}}, \bibinfo
  {author} {\bibfnamefont {C.}~\bibnamefont {Folke}},\ and\ \bibinfo {author}
  {\bibfnamefont {B.}~\bibnamefont {Walker}},\ }\bibfield  {title} {\enquote
  {\bibinfo {title} {Catastrophic shifts in ecosystems},}\ }\href
  {https://doi.org/10.1038/35098000} {\bibfield  {journal} {\bibinfo  {journal}
  {Nature}\ }\textbf {\bibinfo {volume} {413}},\ \bibinfo {pages} {591 – 596}
  (\bibinfo {year} {2001})}\BibitemShut {NoStop}%
\bibitem [{\citenamefont {Bel}, \citenamefont {Hagberg},\ and\ \citenamefont
  {Meron}(2012)}]{VegIsrael2012Bel}%
  \BibitemOpen
  \bibfield  {author} {\bibinfo {author} {\bibfnamefont {G.}~\bibnamefont
  {Bel}}, \bibinfo {author} {\bibfnamefont {A.}~\bibnamefont {Hagberg}},\ and\
  \bibinfo {author} {\bibfnamefont {E.}~\bibnamefont {Meron}},\ }\bibfield
  {title} {\enquote {\bibinfo {title} {Gradual regime shifts in spatially
  extended ecosystems},}\ }\href {https://doi.org/10.1007/s12080-011-0149-6}
  {\bibfield  {journal} {\bibinfo  {journal} {Theoretical Ecology}\ }\textbf
  {\bibinfo {volume} {5}},\ \bibinfo {pages} {591--604} (\bibinfo {year}
  {2012})}\BibitemShut {NoStop}%
\bibitem [{\citenamefont {Franklin}\ \emph {et~al.}(2020)\citenamefont
  {Franklin}, \citenamefont {Harrison}, \citenamefont {Dewar}, \citenamefont
  {Farrior}, \citenamefont {Brännström}, \citenamefont {Dieckmann},
  \citenamefont {Pietsch}, \citenamefont {Falster}, \citenamefont {Cramer},
  \citenamefont {Loreau}, \citenamefont {Wang}, \citenamefont {Mäkelä},
  \citenamefont {Rebel}, \citenamefont {Meron}, \citenamefont {Schymanski},
  \citenamefont {Rovenskaya}, \citenamefont {Stocker}, \citenamefont {Zaehle},
  \citenamefont {Manzoni}, \citenamefont {van Oijen}, \citenamefont {Wright},
  \citenamefont {Ciais}, \citenamefont {van Bodegom}, \citenamefont
  {Peñuelas}, \citenamefont {Hofhansl}, \citenamefont {Terrer}, \citenamefont
  {Soudzilovskaia}, \citenamefont {Midgley},\ and\ \citenamefont
  {Prentice}}]{franklin_organizing_2020}%
  \BibitemOpen
  \bibfield  {author} {\bibinfo {author} {\bibfnamefont {O.}~\bibnamefont
  {Franklin}}, \bibinfo {author} {\bibfnamefont {S.~P.}\ \bibnamefont
  {Harrison}}, \bibinfo {author} {\bibfnamefont {R.}~\bibnamefont {Dewar}},
  \bibinfo {author} {\bibfnamefont {C.~E.}\ \bibnamefont {Farrior}}, \bibinfo
  {author} {\bibfnamefont {k.}~\bibnamefont {Brännström}}, \bibinfo {author}
  {\bibfnamefont {U.}~\bibnamefont {Dieckmann}}, \bibinfo {author}
  {\bibfnamefont {S.}~\bibnamefont {Pietsch}}, \bibinfo {author} {\bibfnamefont
  {D.}~\bibnamefont {Falster}}, \bibinfo {author} {\bibfnamefont
  {W.}~\bibnamefont {Cramer}}, \bibinfo {author} {\bibfnamefont
  {M.}~\bibnamefont {Loreau}}, \bibinfo {author} {\bibfnamefont
  {H.}~\bibnamefont {Wang}}, \bibinfo {author} {\bibfnamefont {A.}~\bibnamefont
  {Mäkelä}}, \bibinfo {author} {\bibfnamefont {K.~T.}\ \bibnamefont {Rebel}},
  \bibinfo {author} {\bibfnamefont {E.}~\bibnamefont {Meron}}, \bibinfo
  {author} {\bibfnamefont {S.~J.}\ \bibnamefont {Schymanski}}, \bibinfo
  {author} {\bibfnamefont {E.}~\bibnamefont {Rovenskaya}}, \bibinfo {author}
  {\bibfnamefont {B.~D.}\ \bibnamefont {Stocker}}, \bibinfo {author}
  {\bibfnamefont {S.}~\bibnamefont {Zaehle}}, \bibinfo {author} {\bibfnamefont
  {S.}~\bibnamefont {Manzoni}}, \bibinfo {author} {\bibfnamefont
  {M.}~\bibnamefont {van Oijen}}, \bibinfo {author} {\bibfnamefont {I.~J.}\
  \bibnamefont {Wright}}, \bibinfo {author} {\bibfnamefont {P.}~\bibnamefont
  {Ciais}}, \bibinfo {author} {\bibfnamefont {P.~M.}\ \bibnamefont {van
  Bodegom}}, \bibinfo {author} {\bibfnamefont {J.}~\bibnamefont {Peñuelas}},
  \bibinfo {author} {\bibfnamefont {F.}~\bibnamefont {Hofhansl}}, \bibinfo
  {author} {\bibfnamefont {C.}~\bibnamefont {Terrer}}, \bibinfo {author}
  {\bibfnamefont {N.~A.}\ \bibnamefont {Soudzilovskaia}}, \bibinfo {author}
  {\bibfnamefont {G.}~\bibnamefont {Midgley}},\ and\ \bibinfo {author}
  {\bibfnamefont {I.~C.}\ \bibnamefont {Prentice}},\ }\bibfield  {title}
  {\enquote {\bibinfo {title} {Organizing principles for vegetation
  dynamics},}\ }\href {https://doi.org/10.1038/s41477-020-0655-x} {\bibfield
  {journal} {\bibinfo  {journal} {Nature Plants}\ }\textbf {\bibinfo {volume}
  {6}},\ \bibinfo {pages} {444--453} (\bibinfo {year} {2020})},\ \bibinfo
  {note} {number: 5 Publisher: Nature Publishing Group}\BibitemShut {NoStop}%
\bibitem [{\citenamefont {Parra-Rivas}\ and\ \citenamefont
  {Fernandez-Oto}(2020)}]{parra-rivas_formation_2020}%
  \BibitemOpen
  \bibfield  {author} {\bibinfo {author} {\bibfnamefont {P.}~\bibnamefont
  {Parra-Rivas}}\ and\ \bibinfo {author} {\bibfnamefont {C.}~\bibnamefont
  {Fernandez-Oto}},\ }\bibfield  {title} {\enquote {\bibinfo {title} {Formation
  of localized states in dryland vegetation: {Bifurcation} structure and
  stability},}\ }\href {https://doi.org/10.1103/PhysRevE.101.052214} {\bibfield
   {journal} {\bibinfo  {journal} {Physical Review E}\ }\textbf {\bibinfo
  {volume} {101}},\ \bibinfo {pages} {052214} (\bibinfo {year}
  {2020})}\BibitemShut {NoStop}%
\bibitem [{\citenamefont {Zelnik}, \citenamefont {Meron},\ and\ \citenamefont
  {Bel}(2015{\natexlab{a}})}]{zelnik_gradual_2015}%
  \BibitemOpen
  \bibfield  {author} {\bibinfo {author} {\bibfnamefont {Y.~R.}\ \bibnamefont
  {Zelnik}}, \bibinfo {author} {\bibfnamefont {E.}~\bibnamefont {Meron}},\ and\
  \bibinfo {author} {\bibfnamefont {G.}~\bibnamefont {Bel}},\ }\bibfield
  {title} {\enquote {\bibinfo {title} {Gradual regime shifts in fairy
  circles},}\ }\href {https://doi.org/10.1073/pnas.1504289112} {\bibfield
  {journal} {\bibinfo  {journal} {Proceedings of the National Academy of
  Sciences}\ }\textbf {\bibinfo {volume} {112}},\ \bibinfo {pages}
  {12327--12331} (\bibinfo {year} {2015}{\natexlab{a}})},\ \bibinfo {note}
  {publisher: Proceedings of the National Academy of Sciences}\BibitemShut
  {NoStop}%
\bibitem [{\citenamefont {Zelnik}\ and\ \citenamefont
  {Tzuk}(2017)}]{zelnik_wavelength_2017}%
  \BibitemOpen
  \bibfield  {author} {\bibinfo {author} {\bibfnamefont {Y.~R.}\ \bibnamefont
  {Zelnik}}\ and\ \bibinfo {author} {\bibfnamefont {O.}~\bibnamefont {Tzuk}},\
  }\bibfield  {title} {\enquote {\bibinfo {title} {Wavelength selection beyond
  turing},}\ }\href {https://doi.org/10.1140/epjst/e2017-70034-x} {\bibfield
  {journal} {\bibinfo  {journal} {The European Physical Journal Special
  Topics}\ }\textbf {\bibinfo {volume} {226}},\ \bibinfo {pages} {2171--2184}
  (\bibinfo {year} {2017})}\BibitemShut {NoStop}%
\bibitem [{\citenamefont {Gilad}\ \emph {et~al.}(2004)\citenamefont {Gilad},
  \citenamefont {von Hardenberg}, \citenamefont {Provenzale}, \citenamefont
  {Shachak},\ and\ \citenamefont {Meron}}]{gilad2004ecosystem}%
  \BibitemOpen
  \bibfield  {author} {\bibinfo {author} {\bibfnamefont {E.}~\bibnamefont
  {Gilad}}, \bibinfo {author} {\bibfnamefont {J.}~\bibnamefont {von
  Hardenberg}}, \bibinfo {author} {\bibfnamefont {A.}~\bibnamefont
  {Provenzale}}, \bibinfo {author} {\bibfnamefont {M.}~\bibnamefont
  {Shachak}},\ and\ \bibinfo {author} {\bibfnamefont {E.}~\bibnamefont
  {Meron}},\ }\bibfield  {title} {\enquote {\bibinfo {title} {Ecosystem
  engineers: from pattern formation to habitat creation},}\ }\href@noop {}
  {\bibfield  {journal} {\bibinfo  {journal} {Physical Review Letters}\
  }\textbf {\bibinfo {volume} {93}},\ \bibinfo {pages} {098105} (\bibinfo
  {year} {2004})}\BibitemShut {NoStop}%
\bibitem [{\citenamefont {Meron}(2016)}]{meron2016pattern}%
  \BibitemOpen
  \bibfield  {author} {\bibinfo {author} {\bibfnamefont {E.}~\bibnamefont
  {Meron}},\ }\bibfield  {title} {\enquote {\bibinfo {title} {Pattern
  formation--a missing link in the study of ecosystem response to environmental
  changes},}\ }\href@noop {} {\bibfield  {journal} {\bibinfo  {journal}
  {Mathematical biosciences}\ }\textbf {\bibinfo {volume} {271}},\ \bibinfo
  {pages} {1--18} (\bibinfo {year} {2016})}\BibitemShut {NoStop}%
\bibitem [{\citenamefont {Zelnik}, \citenamefont {Meron},\ and\ \citenamefont
  {Bel}(2015{\natexlab{b}})}]{zelnik2015gradual}%
  \BibitemOpen
  \bibfield  {author} {\bibinfo {author} {\bibfnamefont {Y.~R.}\ \bibnamefont
  {Zelnik}}, \bibinfo {author} {\bibfnamefont {E.}~\bibnamefont {Meron}},\ and\
  \bibinfo {author} {\bibfnamefont {G.}~\bibnamefont {Bel}},\ }\bibfield
  {title} {\enquote {\bibinfo {title} {Gradual regime shifts in fairy
  circles},}\ }\href@noop {} {\bibfield  {journal} {\bibinfo  {journal}
  {Proceedings of the National Academy of Sciences}\ }\textbf {\bibinfo
  {volume} {112}},\ \bibinfo {pages} {12327--12331} (\bibinfo {year}
  {2015}{\natexlab{b}})}\BibitemShut {NoStop}%
\bibitem [{\citenamefont {Fernandez-Oto}, \citenamefont {Tzuk},\ and\
  \citenamefont {Meron}(2019)}]{fernandez2019front}%
  \BibitemOpen
  \bibfield  {author} {\bibinfo {author} {\bibfnamefont {C.}~\bibnamefont
  {Fernandez-Oto}}, \bibinfo {author} {\bibfnamefont {O.}~\bibnamefont
  {Tzuk}},\ and\ \bibinfo {author} {\bibfnamefont {E.}~\bibnamefont {Meron}},\
  }\bibfield  {title} {\enquote {\bibinfo {title} {Front instabilities can
  reverse desertification},}\ }\href@noop {} {\bibfield  {journal} {\bibinfo
  {journal} {Physical review letters}\ }\textbf {\bibinfo {volume} {122}},\
  \bibinfo {pages} {048101} (\bibinfo {year} {2019})}\BibitemShut {NoStop}%
\bibitem [{\citenamefont {Meixner}\ \emph {et~al.}(1997)\citenamefont
  {Meixner}, \citenamefont {De~Wit}, \citenamefont {Bose},\ and\ \citenamefont
  {Schöll}}]{meixner_generic_1997}%
  \BibitemOpen
  \bibfield  {author} {\bibinfo {author} {\bibfnamefont {M.}~\bibnamefont
  {Meixner}}, \bibinfo {author} {\bibfnamefont {A.}~\bibnamefont {De~Wit}},
  \bibinfo {author} {\bibfnamefont {S.}~\bibnamefont {Bose}},\ and\ \bibinfo
  {author} {\bibfnamefont {E.}~\bibnamefont {Schöll}},\ }\bibfield  {title}
  {\enquote {\bibinfo {title} {Generic spatiotemporal dynamics near
  codimension-two {Turing}-{Hopf} bifurcations},}\ }\href
  {https://doi.org/10.1103/PhysRevE.55.6690} {\bibfield  {journal} {\bibinfo
  {journal} {Physical Review E}\ }\textbf {\bibinfo {volume} {55}},\ \bibinfo
  {pages} {6690--6697} (\bibinfo {year} {1997})},\ \bibinfo {note} {publisher:
  American Physical Society}\BibitemShut {NoStop}%
\bibitem [{\citenamefont {Just}\ \emph {et~al.}(2001)\citenamefont {Just},
  \citenamefont {Bose}, \citenamefont {Bose}, \citenamefont {Engel},\ and\
  \citenamefont {Schöll}}]{just_spatiotemporal_2001}%
  \BibitemOpen
  \bibfield  {author} {\bibinfo {author} {\bibfnamefont {W.}~\bibnamefont
  {Just}}, \bibinfo {author} {\bibfnamefont {M.}~\bibnamefont {Bose}}, \bibinfo
  {author} {\bibfnamefont {S.}~\bibnamefont {Bose}}, \bibinfo {author}
  {\bibfnamefont {H.}~\bibnamefont {Engel}},\ and\ \bibinfo {author}
  {\bibfnamefont {E.}~\bibnamefont {Schöll}},\ }\bibfield  {title} {\enquote
  {\bibinfo {title} {Spatiotemporal dynamics near a supercritical
  {Turing}-{Hopf} bifurcation in a two-dimensional reaction-diffusion
  system},}\ }\href {https://doi.org/10.1103/PhysRevE.64.026219} {\bibfield
  {journal} {\bibinfo  {journal} {Physical Review E}\ }\textbf {\bibinfo
  {volume} {64}},\ \bibinfo {pages} {026219} (\bibinfo {year} {2001})},\
  \bibinfo {note} {publisher: American Physical Society}\BibitemShut {NoStop}%
\bibitem [{\citenamefont {Allgower}\ and\ \citenamefont
  {Georg}(1990)}]{allgower_numerical_1990}%
  \BibitemOpen
  \bibfield  {author} {\bibinfo {author} {\bibfnamefont {E.~L.}\ \bibnamefont
  {Allgower}}\ and\ \bibinfo {author} {\bibfnamefont {K.}~\bibnamefont
  {Georg}},\ }\href {https://www.springer.com/gp/book/9783642647642} {\emph
  {\bibinfo {title} {Numerical {Continuation} {Methods}: {An}
  {Introduction}}}},\ Springer {Series} in {Computational} {Mathematics}\
  (\bibinfo  {publisher} {Springer-Verlag},\ \bibinfo {address} {Berlin
  Heidelberg},\ \bibinfo {year} {1990})\BibitemShut {NoStop}%
\bibitem [{\citenamefont {Krauskopf}, \citenamefont {Osinga},\ and\
  \citenamefont {Galan-Vioque}(2007)}]{krauskopf_numerical_2007}%
  \BibitemOpen
  \bibinfo {editor} {\bibfnamefont {B.}~\bibnamefont {Krauskopf}}, \bibinfo
  {editor} {\bibfnamefont {H.~M.}\ \bibnamefont {Osinga}},\ and\ \bibinfo
  {editor} {\bibfnamefont {J.}~\bibnamefont {Galan-Vioque}},\ eds.,\ \href
  {https://doi.org/10.1007/978-1-4020-6356-5} {\emph {\bibinfo {title}
  {Numerical {Continuation} {Methods} for {Dynamical} {Systems}: {Path}
  following and boundary value problems}}},\ Understanding {Complex} {Systems}\
  (\bibinfo  {publisher} {Springer Netherlands},\ \bibinfo {year}
  {2007})\BibitemShut {NoStop}%
\bibitem [{\citenamefont {Uecker}(2021)}]{uecker_numerical_2021}%
  \BibitemOpen
  \bibfield  {author} {\bibinfo {author} {\bibfnamefont {H.}~\bibnamefont
  {Uecker}},\ }\href {https://doi.org/10.1137/1.9781611976618} {\emph {\bibinfo
  {title} {Numerical {Continuation} and {Bifurcation} in {Nonlinear}
  {PDEs}}}},\ Other {Titles} in {Applied} {Mathematics}\ (\bibinfo  {publisher}
  {Society for Industrial and Applied Mathematics},\ \bibinfo {year}
  {2021})\BibitemShut {NoStop}%
\bibitem [{\citenamefont {Doedel}\ \emph {et~al.}(2007)\citenamefont {Doedel},
  \citenamefont {Fairgrieve}, \citenamefont {Sandstede}, \citenamefont
  {Champneys}, \citenamefont {Kuznetsov},\ and\ \citenamefont
  {Wang}}]{doedel_auto-07p_2007}%
  \BibitemOpen
  \bibfield  {author} {\bibinfo {author} {\bibfnamefont {E.~J.}\ \bibnamefont
  {Doedel}}, \bibinfo {author} {\bibfnamefont {T.~F.}\ \bibnamefont
  {Fairgrieve}}, \bibinfo {author} {\bibfnamefont {B.}~\bibnamefont
  {Sandstede}}, \bibinfo {author} {\bibfnamefont {A.~R.}\ \bibnamefont
  {Champneys}}, \bibinfo {author} {\bibfnamefont {Y.~A.}\ \bibnamefont
  {Kuznetsov}},\ and\ \bibinfo {author} {\bibfnamefont {X.}~\bibnamefont
  {Wang}},\ }\href@noop {} {\enquote {\bibinfo {title} {{AUTO}-{07P}:
  {Continuation} and bifurcation software for ordinary differential
  equations},}\ }\bibinfo {type} {Tech. Rep.}\ (\bibinfo {year}
  {2007})\BibitemShut {NoStop}%
\bibitem [{\citenamefont {Colet}\ \emph {et~al.}(2014)\citenamefont {Colet},
  \citenamefont {Mat\'ias}, \citenamefont {Gelens},\ and\ \citenamefont
  {Gomila}}]{colet_formation_2014}%
  \BibitemOpen
  \bibfield  {author} {\bibinfo {author} {\bibfnamefont {P.}~\bibnamefont
  {Colet}}, \bibinfo {author} {\bibfnamefont {M.~A.}\ \bibnamefont {Mat\'ias}},
  \bibinfo {author} {\bibfnamefont {L.}~\bibnamefont {Gelens}},\ and\ \bibinfo
  {author} {\bibfnamefont {D.}~\bibnamefont {Gomila}},\ }\bibfield  {title}
  {\enquote {\bibinfo {title} {Formation of localized structures in bistable
  systems through nonlocal spatial coupling. {I}. {General} framework},}\
  }\href {https://doi.org/10.1103/PhysRevE.89.012914} {\bibfield  {journal}
  {\bibinfo  {journal} {Physical Review E}\ }\textbf {\bibinfo {volume} {89}},\
  \bibinfo {pages} {012914} (\bibinfo {year} {2014})}\BibitemShut {NoStop}%
\bibitem [{\citenamefont {Woods}\ and\ \citenamefont
  {Champneys}(1999)}]{woods_heteroclinic_1999}%
  \BibitemOpen
  \bibfield  {author} {\bibinfo {author} {\bibfnamefont {P.~D.}\ \bibnamefont
  {Woods}}\ and\ \bibinfo {author} {\bibfnamefont {A.~R.}\ \bibnamefont
  {Champneys}},\ }\bibfield  {title} {\enquote {\bibinfo {title} {Heteroclinic
  tangles and homoclinic snaking in the unfolding of a degenerate reversible
  {Hamiltonian}–{Hopf} bifurcation},}\ }\href
  {https://doi.org/10.1016/S0167-2789(98)00309-1} {\bibfield  {journal}
  {\bibinfo  {journal} {Physica D: Nonlinear Phenomena}\ }\textbf {\bibinfo
  {volume} {129}},\ \bibinfo {pages} {147--170} (\bibinfo {year}
  {1999})}\BibitemShut {NoStop}%
\bibitem [{\citenamefont {Coullet}, \citenamefont {Riera},\ and\ \citenamefont
  {Tresser}(2000)}]{coullet_stable_2000}%
  \BibitemOpen
  \bibfield  {author} {\bibinfo {author} {\bibfnamefont {P.}~\bibnamefont
  {Coullet}}, \bibinfo {author} {\bibfnamefont {C.}~\bibnamefont {Riera}},\
  and\ \bibinfo {author} {\bibfnamefont {C.}~\bibnamefont {Tresser}},\
  }\bibfield  {title} {\enquote {\bibinfo {title} {Stable {Static} {Localized}
  {Structures} in {One} {Dimension}},}\ }\href
  {https://doi.org/10.1103/PhysRevLett.84.3069} {\bibfield  {journal} {\bibinfo
   {journal} {Physical Review Letters}\ }\textbf {\bibinfo {volume} {84}},\
  \bibinfo {pages} {3069--3072} (\bibinfo {year} {2000})}\BibitemShut {NoStop}%
\bibitem [{\citenamefont {Makrides}\ and\ \citenamefont
  {Sandstede}(2019)}]{makrides_existence_2019}%
  \BibitemOpen
  \bibfield  {author} {\bibinfo {author} {\bibfnamefont {E.}~\bibnamefont
  {Makrides}}\ and\ \bibinfo {author} {\bibfnamefont {B.}~\bibnamefont
  {Sandstede}},\ }\bibfield  {title} {\enquote {\bibinfo {title} {Existence and
  stability of spatially localized patterns},}\ }\href
  {https://doi.org/10.1016/j.jde.2018.07.064} {\bibfield  {journal} {\bibinfo
  {journal} {Journal of Differential Equations}\ }\textbf {\bibinfo {volume}
  {266}},\ \bibinfo {pages} {1073--1120} (\bibinfo {year} {2019})}\BibitemShut
  {NoStop}%
\bibitem [{\citenamefont {Knobloch}\ and\ \citenamefont
  {Wagenknecht}(2005)}]{knobloch_homoclinic_2005}%
  \BibitemOpen
  \bibfield  {author} {\bibinfo {author} {\bibfnamefont {J.}~\bibnamefont
  {Knobloch}}\ and\ \bibinfo {author} {\bibfnamefont {T.}~\bibnamefont
  {Wagenknecht}},\ }\bibfield  {title} {\enquote {\bibinfo {title} {Homoclinic
  snaking near a heteroclinic cycle in reversible systems},}\ }\href
  {https://doi.org/10.1016/j.physd.2005.04.018} {\bibfield  {journal} {\bibinfo
   {journal} {Physica D: Nonlinear Phenomena}\ }\textbf {\bibinfo {volume}
  {206}},\ \bibinfo {pages} {82--93} (\bibinfo {year} {2005})}\BibitemShut
  {NoStop}%
\bibitem [{\citenamefont {Zelnik}\ \emph {et~al.}(2018)\citenamefont {Zelnik},
  \citenamefont {Gandhi}, \citenamefont {Knobloch},\ and\ \citenamefont
  {Meron}}]{zelnik_implications_2018}%
  \BibitemOpen
  \bibfield  {author} {\bibinfo {author} {\bibfnamefont {Y.~R.}\ \bibnamefont
  {Zelnik}}, \bibinfo {author} {\bibfnamefont {P.}~\bibnamefont {Gandhi}},
  \bibinfo {author} {\bibfnamefont {E.}~\bibnamefont {Knobloch}},\ and\
  \bibinfo {author} {\bibfnamefont {E.}~\bibnamefont {Meron}},\ }\bibfield
  {title} {\enquote {\bibinfo {title} {Implications of tristability in
  pattern-forming ecosystems},}\ }\href {https://doi.org/10.1063/1.5018925}
  {\bibfield  {journal} {\bibinfo  {journal} {Chaos: An Interdisciplinary
  Journal of Nonlinear Science}\ }\textbf {\bibinfo {volume} {28}},\ \bibinfo
  {pages} {033609} (\bibinfo {year} {2018})}\BibitemShut {NoStop}%
\bibitem [{\citenamefont {Parra-Rivas}\ \emph {et~al.}(2022)\citenamefont
  {Parra-Rivas}, \citenamefont {Champneys}, \citenamefont {Al-Sahadi},
  \citenamefont {Gomila},\ and\ \citenamefont
  {Knobloch}}]{parra-rivas_organization_2022}%
  \BibitemOpen
  \bibfield  {author} {\bibinfo {author} {\bibfnamefont {P.}~\bibnamefont
  {Parra-Rivas}}, \bibinfo {author} {\bibfnamefont {A.~R.}\ \bibnamefont
  {Champneys}}, \bibinfo {author} {\bibfnamefont {F.}~\bibnamefont
  {Al-Sahadi}}, \bibinfo {author} {\bibfnamefont {D.}~\bibnamefont {Gomila}},\
  and\ \bibinfo {author} {\bibfnamefont {E.}~\bibnamefont {Knobloch}},\ }\href
  {https://doi.org/10.48550/arXiv.2208.04009} {\enquote {\bibinfo {title}
  {Organization of spatially localized structures near a codimension-three
  cusp-{Turing} bifurcation},}\ } (\bibinfo {year} {2022}),\ \bibinfo {note}
  {arXiv:2208.04009 [nlin]}\BibitemShut {NoStop}%
\bibitem [{\citenamefont {Burke}\ and\ \citenamefont
  {Knobloch}(2007{\natexlab{a}})}]{burke_homoclinic_2007}%
  \BibitemOpen
  \bibfield  {author} {\bibinfo {author} {\bibfnamefont {J.}~\bibnamefont
  {Burke}}\ and\ \bibinfo {author} {\bibfnamefont {E.}~\bibnamefont
  {Knobloch}},\ }\bibfield  {title} {\enquote {\bibinfo {title} {Homoclinic
  snaking: {Structure} and stability},}\ }\href
  {https://doi.org/10.1063/1.2746816} {\bibfield  {journal} {\bibinfo
  {journal} {Chaos: An Interdisciplinary Journal of Nonlinear Science}\
  }\textbf {\bibinfo {volume} {17}},\ \bibinfo {pages} {037102} (\bibinfo
  {year} {2007}{\natexlab{a}})}\BibitemShut {NoStop}%
\bibitem [{\citenamefont {Burke}\ and\ \citenamefont
  {Knobloch}(2007{\natexlab{b}})}]{burke_snakes_2007}%
  \BibitemOpen
  \bibfield  {author} {\bibinfo {author} {\bibfnamefont {J.}~\bibnamefont
  {Burke}}\ and\ \bibinfo {author} {\bibfnamefont {E.}~\bibnamefont
  {Knobloch}},\ }\bibfield  {title} {\enquote {\bibinfo {title} {Snakes and
  ladders: {Localized} states in the {Swift}–{Hohenberg} equation},}\ }\href
  {https://doi.org/10.1016/j.physleta.2006.08.072} {\bibfield  {journal}
  {\bibinfo  {journal} {Physics Letters A}\ }\textbf {\bibinfo {volume}
  {360}},\ \bibinfo {pages} {681--688} (\bibinfo {year}
  {2007}{\natexlab{b}})}\BibitemShut {NoStop}%
\bibitem [{\citenamefont {Parra-Rivas}\ \emph {et~al.}(2018)\citenamefont
  {Parra-Rivas}, \citenamefont {Gomila}, \citenamefont {Gelens},\ and\
  \citenamefont {Knobloch}}]{parra-rivas_bifurcation_2018}%
  \BibitemOpen
  \bibfield  {author} {\bibinfo {author} {\bibfnamefont {P.}~\bibnamefont
  {Parra-Rivas}}, \bibinfo {author} {\bibfnamefont {D.}~\bibnamefont {Gomila}},
  \bibinfo {author} {\bibfnamefont {L.}~\bibnamefont {Gelens}},\ and\ \bibinfo
  {author} {\bibfnamefont {E.}~\bibnamefont {Knobloch}},\ }\bibfield  {title}
  {\enquote {\bibinfo {title} {Bifurcation structure of localized states in the
  {Lugiato}-{Lefever} equation with anomalous dispersion},}\ }\href
  {https://doi.org/10.1103/PhysRevE.97.042204} {\bibfield  {journal} {\bibinfo
  {journal} {Physical Review E}\ }\textbf {\bibinfo {volume} {97}},\ \bibinfo
  {pages} {042204} (\bibinfo {year} {2018})}\BibitemShut {NoStop}%
\bibitem [{\citenamefont {Al~Saadi}\ and\ \citenamefont
  {Champneys}(2021)}]{al_saadi_unified_2021}%
  \BibitemOpen
  \bibfield  {author} {\bibinfo {author} {\bibfnamefont {F.}~\bibnamefont
  {Al~Saadi}}\ and\ \bibinfo {author} {\bibfnamefont {A.}~\bibnamefont
  {Champneys}},\ }\bibfield  {title} {\enquote {\bibinfo {title} {Unified
  framework for localized patterns in reaction–diffusion systems; the
  {Gray}–{Scott} and {Gierer}–{Meinhardt} cases},}\ }\href
  {https://doi.org/10.1098/rsta.2020.0277} {\bibfield  {journal} {\bibinfo
  {journal} {Philosophical Transactions of the Royal Society A: Mathematical,
  Physical and Engineering Sciences}\ }\textbf {\bibinfo {volume} {379}},\
  \bibinfo {pages} {20200277} (\bibinfo {year} {2021})},\ \bibinfo {note}
  {publisher: Royal Society}\BibitemShut {NoStop}%
\bibitem [{\citenamefont {{Al Saadi}}\ \emph {et~al.}(2022)\citenamefont {{Al
  Saadi}}, \citenamefont {Worthy}, \citenamefont {Alrihieli},\ and\
  \citenamefont {Nelson}}]{ALSAADI2022}%
  \BibitemOpen
  \bibfield  {author} {\bibinfo {author} {\bibfnamefont {F.}~\bibnamefont {{Al
  Saadi}}}, \bibinfo {author} {\bibfnamefont {A.}~\bibnamefont {Worthy}},
  \bibinfo {author} {\bibfnamefont {H.}~\bibnamefont {Alrihieli}},\ and\
  \bibinfo {author} {\bibfnamefont {M.}~\bibnamefont {Nelson}},\ }\bibfield
  {title} {\enquote {\bibinfo {title} {Localised spatial structures in the
  thomas model},}\ }\href@noop {} {\bibfield  {journal} {\bibinfo  {journal}
  {Mathematics and Computers in Simulation}\ }\textbf {\bibinfo {volume}
  {194}},\ \bibinfo {pages} {141--158} (\bibinfo {year} {2022})}\BibitemShut
  {NoStop}%
\bibitem [{\citenamefont {Parra-Rivas}\ \emph {et~al.}(2021)\citenamefont
  {Parra-Rivas}, \citenamefont {Knobloch}, \citenamefont {Gelens},\ and\
  \citenamefont {Gomila}}]{parra-rivas_origin_2021}%
  \BibitemOpen
  \bibfield  {author} {\bibinfo {author} {\bibfnamefont {P.}~\bibnamefont
  {Parra-Rivas}}, \bibinfo {author} {\bibfnamefont {E.}~\bibnamefont
  {Knobloch}}, \bibinfo {author} {\bibfnamefont {L.}~\bibnamefont {Gelens}},\
  and\ \bibinfo {author} {\bibfnamefont {D.}~\bibnamefont {Gomila}},\
  }\bibfield  {title} {\enquote {\bibinfo {title} {Origin, bifurcation
  structure and stability of localized states in {Kerr} dispersive optical
  cavities},}\ }\href {https://doi.org/10.1093/imamat/hxab031} {\bibfield
  {journal} {\bibinfo  {journal} {IMA Journal of Applied Mathematics}\ }\textbf
  {\bibinfo {volume} {86}},\ \bibinfo {pages} {856--895} (\bibinfo {year}
  {2021})}\BibitemShut {NoStop}%
\bibitem [{\citenamefont {Borckmans}\ \emph {et~al.}(1995)\citenamefont
  {Borckmans}, \citenamefont {Jensen}, \citenamefont {Pannbacker},
  \citenamefont {Mosekilde}, \citenamefont {Dewel},\ and\ \citenamefont
  {De~Wit}}]{borckmans_localized_1995}%
  \BibitemOpen
  \bibfield  {author} {\bibinfo {author} {\bibfnamefont {P.}~\bibnamefont
  {Borckmans}}, \bibinfo {author} {\bibfnamefont {O.}~\bibnamefont {Jensen}},
  \bibinfo {author} {\bibfnamefont {V.~O.}\ \bibnamefont {Pannbacker}},
  \bibinfo {author} {\bibfnamefont {E.}~\bibnamefont {Mosekilde}}, \bibinfo
  {author} {\bibfnamefont {G.}~\bibnamefont {Dewel}},\ and\ \bibinfo {author}
  {\bibfnamefont {A.}~\bibnamefont {De~Wit}},\ }\bibfield  {title} {\enquote
  {\bibinfo {title} {Localized {Turing} and {Turing}-{Hopf} {Patterns}},}\ }in\
  \href {https://doi.org/10.1007/978-3-642-79290-8_4} {\emph {\bibinfo
  {booktitle} {Modelling the {Dynamics} of {Biological} {Systems}: {Nonlinear}
  {Phenomena} and {Pattern} {Formation}}}},\ \bibinfo {series and number}
  {Springer {Series} in {Synergetics}},\ \bibinfo {editor} {edited by\ \bibinfo
  {editor} {\bibfnamefont {E.}~\bibnamefont {Mosekilde}}\ and\ \bibinfo
  {editor} {\bibfnamefont {O.~G.}\ \bibnamefont {Mouritsen}}}\ (\bibinfo
  {publisher} {Springer},\ \bibinfo {address} {Berlin, Heidelberg},\ \bibinfo
  {year} {1995})\ pp.\ \bibinfo {pages} {48--73}\BibitemShut {NoStop}%
\bibitem [{\citenamefont {De~Wit}\ \emph {et~al.}(1996)\citenamefont {De~Wit},
  \citenamefont {Lima}, \citenamefont {Dewel},\ and\ \citenamefont
  {Borckmans}}]{de_wit_spatiotemporal_1996}%
  \BibitemOpen
  \bibfield  {author} {\bibinfo {author} {\bibfnamefont {A.}~\bibnamefont
  {De~Wit}}, \bibinfo {author} {\bibfnamefont {D.}~\bibnamefont {Lima}},
  \bibinfo {author} {\bibfnamefont {G.}~\bibnamefont {Dewel}},\ and\ \bibinfo
  {author} {\bibfnamefont {P.}~\bibnamefont {Borckmans}},\ }\bibfield  {title}
  {\enquote {\bibinfo {title} {Spatiotemporal dynamics near a codimension-two
  point},}\ }\href {https://doi.org/10.1103/PhysRevE.54.261} {\bibfield
  {journal} {\bibinfo  {journal} {Physical Review E}\ }\textbf {\bibinfo
  {volume} {54}},\ \bibinfo {pages} {261--271} (\bibinfo {year}
  {1996})}\BibitemShut {NoStop}%
\bibitem [{\citenamefont {Tlidi}, \citenamefont {Mandel},\ and\ \citenamefont
  {Haelterman}(1997)}]{tlidi_spatiotemporal_1997}%
  \BibitemOpen
  \bibfield  {author} {\bibinfo {author} {\bibfnamefont {M.}~\bibnamefont
  {Tlidi}}, \bibinfo {author} {\bibfnamefont {P.}~\bibnamefont {Mandel}},\ and\
  \bibinfo {author} {\bibfnamefont {M.}~\bibnamefont {Haelterman}},\ }\bibfield
   {title} {\enquote {\bibinfo {title} {Spatiotemporal patterns and localized
  structures in nonlinear optics},}\ }\href
  {https://doi.org/10.1103/PhysRevE.56.6524} {\bibfield  {journal} {\bibinfo
  {journal} {Physical Review E}\ }\textbf {\bibinfo {volume} {56}},\ \bibinfo
  {pages} {6524--6530} (\bibinfo {year} {1997})},\ \bibinfo {note} {publisher:
  American Physical Society}\BibitemShut {NoStop}%
\bibitem [{\citenamefont {Tzou}\ \emph {et~al.}(2013)\citenamefont {Tzou},
  \citenamefont {Ma}, \citenamefont {Bayliss}, \citenamefont {Matkowsky},\ and\
  \citenamefont {Volpert}}]{tzou_homoclinic_2013}%
  \BibitemOpen
  \bibfield  {author} {\bibinfo {author} {\bibfnamefont {J.~C.}\ \bibnamefont
  {Tzou}}, \bibinfo {author} {\bibfnamefont {Y.-P.}\ \bibnamefont {Ma}},
  \bibinfo {author} {\bibfnamefont {A.}~\bibnamefont {Bayliss}}, \bibinfo
  {author} {\bibfnamefont {B.~J.}\ \bibnamefont {Matkowsky}},\ and\ \bibinfo
  {author} {\bibfnamefont {V.~A.}\ \bibnamefont {Volpert}},\ }\bibfield
  {title} {\enquote {\bibinfo {title} {Homoclinic snaking near a
  codimension-two {Turing}-{Hopf} bifurcation point in the {Brusselator}
  model},}\ }\href {https://doi.org/10.1103/PhysRevE.87.022908} {\bibfield
  {journal} {\bibinfo  {journal} {Physical Review E}\ }\textbf {\bibinfo
  {volume} {87}},\ \bibinfo {pages} {022908} (\bibinfo {year} {2013})},\
  \bibinfo {note} {publisher: American Physical Society}\BibitemShut {NoStop}%
\bibitem [{\citenamefont {Heidemann}, \citenamefont {Bode},\ and\ \citenamefont
  {Purwins}(1993)}]{heidemann_fronts_1993}%
  \BibitemOpen
  \bibfield  {author} {\bibinfo {author} {\bibfnamefont {G.}~\bibnamefont
  {Heidemann}}, \bibinfo {author} {\bibfnamefont {M.}~\bibnamefont {Bode}},\
  and\ \bibinfo {author} {\bibfnamefont {H.~G.}\ \bibnamefont {Purwins}},\
  }\bibfield  {title} {\enquote {\bibinfo {title} {Fronts between {Hopf}- and
  {Turing}-type domains in a two-component reaction-diffusion system},}\ }\href
  {https://doi.org/10.1016/0375-9601(93)90030-4} {\bibfield  {journal}
  {\bibinfo  {journal} {Physics Letters A}\ }\textbf {\bibinfo {volume}
  {177}},\ \bibinfo {pages} {225--230} (\bibinfo {year} {1993})}\BibitemShut
  {NoStop}%
\bibitem [{\citenamefont {Kolodner}(1993)}]{kolodner_coexisting_1993}%
  \BibitemOpen
  \bibfield  {author} {\bibinfo {author} {\bibfnamefont {P.}~\bibnamefont
  {Kolodner}},\ }\bibfield  {title} {\enquote {\bibinfo {title} {Coexisting
  traveling waves and steady rolls in binary-fluid convection},}\ }\href
  {https://doi.org/10.1103/PhysRevE.48.R665} {\bibfield  {journal} {\bibinfo
  {journal} {Physical Review E}\ }\textbf {\bibinfo {volume} {48}},\ \bibinfo
  {pages} {R665--R668} (\bibinfo {year} {1993})},\ \bibinfo {note} {publisher:
  American Physical Society}\BibitemShut {NoStop}%
\bibitem [{\citenamefont {Uecker}, \citenamefont {Wetzel},\ and\ \citenamefont
  {Rademacher}(2014)}]{uecker_pde2path_2014}%
  \BibitemOpen
  \bibfield  {author} {\bibinfo {author} {\bibfnamefont {H.}~\bibnamefont
  {Uecker}}, \bibinfo {author} {\bibfnamefont {D.}~\bibnamefont {Wetzel}},\
  and\ \bibinfo {author} {\bibfnamefont {J.~D.~M.}\ \bibnamefont
  {Rademacher}},\ }\bibfield  {title} {\enquote {\bibinfo {title} {pde2path -
  {A} {Matlab} {Package} for {Continuation} and {Bifurcation} in {2D}
  {Elliptic} {Systems}},}\ }\href {https://doi.org/10.1017/S1004897900000295}
  {\bibfield  {journal} {\bibinfo  {journal} {Numerical Mathematics: Theory,
  Methods and Applications}\ }\textbf {\bibinfo {volume} {7}},\ \bibinfo
  {pages} {58--106} (\bibinfo {year} {2014})}\BibitemShut {NoStop}%
\bibitem [{\citenamefont {Parra-Rivas}, \citenamefont {Mas-Arabí},\ and\
  \citenamefont {Leo}(2020)}]{parra-rivas_parametric_2020}%
  \BibitemOpen
  \bibfield  {author} {\bibinfo {author} {\bibfnamefont {P.}~\bibnamefont
  {Parra-Rivas}}, \bibinfo {author} {\bibfnamefont {C.}~\bibnamefont
  {Mas-Arabí}},\ and\ \bibinfo {author} {\bibfnamefont {F.}~\bibnamefont
  {Leo}},\ }\bibfield  {title} {\enquote {\bibinfo {title} {Parametric
  localized patterns and breathers in dispersive quadratic cavities},}\ }\href
  {http://arxiv.org/abs/2003.09941} {\bibfield  {journal} {\bibinfo  {journal}
  {arXiv:2003.09941 [nlin, physics:physics]}\ } (\bibinfo {year} {2020})},\
  \bibinfo {note} {arXiv: 2003.09941}\BibitemShut {NoStop}%
\end{thebibliography}%


\end{document}